\documentclass[pra,twocolumn,showkeys,superscriptaddress,amsmath,showpacs,tightenlines,longbibliography]{revtex4-1}
\usepackage{dcolumn}
\usepackage{graphicx}
\usepackage{mathrsfs}
\usepackage{subfigure}
\usepackage{booktabs}
\usepackage{multirow}
\usepackage{amsmath}
\usepackage{dsfont}
\usepackage{amstext}
\usepackage{amssymb}
\usepackage{amsbsy}
\usepackage{appendix}
\usepackage{threeparttable}
\usepackage{bbm}
\usepackage{amsthm}
\usepackage{ulem}
\usepackage{graphicx}
\usepackage{mdwlist}
\usepackage{lipsum}
\usepackage{xcolor}
\usepackage[none]{hyphenat}
\usepackage[colorlinks,citecolor=blue]{hyperref}
\definecolor{Dred}{RGB}{240 0 0}
\definecolor{purple}{RGB}{160 32 240}
\usepackage{url}
\usepackage[colorlinks]{hyperref}
\hypersetup{%
	plainpages=true,
	breaklinks=true,  %not default in dvips mode, so we must specify
	hypertexnames=false,  %not ideal, but needed when pagenums duplicate (`i' vs. `1')
	pageanchor=true,
	colorlinks=true,
	linkcolor={blue},
	citecolor={red},
	urlcolor={blue},
	anchorcolor={black}
}

\usepackage{CJKutf8}

\begin{document}
%\begin{CJK*}{UTF8}{gbsn}		
		
		\title{{Fast binomial-code holonomic quantum computation with ultrastrong light-matter coupling}}
    	\author{Ye-Hong Chen}
    	\affiliation{Theoretical Quantum Physics Laboratory, RIKEN Cluster for Pioneering Research, Wako-shi, Saitama 351-0198, Japan}
		
		\author{Wei Qin}
		\email[E-mail: ]{wei.qin@riken.jp}
		\affiliation{Theoretical Quantum Physics Laboratory, RIKEN Cluster for Pioneering Research, Wako-shi, Saitama 351-0198, Japan}
		
		\author{Roberto Stassi}
		\affiliation{Theoretical Quantum Physics Laboratory, RIKEN Cluster for Pioneering Research, Wako-shi, Saitama 351-0198, Japan}
		\affiliation{Dipartimento di Scienze Matematiche e Informatiche, Scienze Fisiche e Scienze della Terra, Universit\`{a} di Messina, 98166, Messina, Italy}

		\author{Xin Wang}
		\affiliation{Theoretical Quantum Physics Laboratory, RIKEN Cluster for Pioneering Research, Wako-shi, Saitama 351-0198, Japan}
		\affiliation{Institute of Quantum Optics and Quantum Information, School of Science, Xi'an Jiaotong University, Xi'an 710049, China}
		
		\author{Franco Nori}
		\email[E-mail: ]{fnori@riken.jp}
		\affiliation{Theoretical Quantum Physics Laboratory, RIKEN Cluster for Pioneering Research, Wako-shi, Saitama 351-0198, Japan}
		\affiliation{Department of Physics, University of Michigan, Ann Arbor, Michigan 48109-1040, USA}
        \affiliation{RIKEN Center for Quantum Computing (RQC), Wako-shi, Saitama 351-0198, Japan}
        
\date{\today}

\begin{abstract}
	{We propose a protocol for bosonic binomial-code nonadiabatic holonomic quantum computation in a system composed of an artificial atom ultrastrongly coupled to a cavity resonator. In our protocol,
    the binomial codes, formed by superpositions of Fock states,
    can greatly save physical resources to correct errors in quantum computation.
    We apply to the system strong driving fields designed by shortcuts-to-adiabatic methods. This reduces the gate time to
    \textit{tens of nanoseconds}.
    %Decoherence of the system accumulated over time can be effectively reduced by such a fast evolution.
    Noise induced by control imperfections
    can be suppressed by a systematic-error-sensitivity nullification
    method. %thus the protocol is mostly insensitive to such noises.
    As a result, this protocol can rapidly ($\sim 35~{\rm{ns}}$) generate fault-tolerant and high-fidelity ($\gtrsim 98\%$ with experimentally realistic parameters) quantum gates.} % in the pesence of decoherence and noises.}
    %Therefore, the proposed protocols can rapidly generate the Fock-state %superpositions and the fault-tolerant holonomic gates with high fidelities.
\end{abstract}

%\pacs {03.67.}

\maketitle

\section{Introduction}

The generation of robust and fault-tolerant quantum gates
is a basic requirement for quantum computation. To reach this goal, much attention
has been given to holonomic quantum computation \cite{Pla26494,Prl89097902,Njp14103035,Prl109170501}
based on Abelian \cite{Prsa39245,Prl581593} and non-Abelian geometric phases \cite{Pla133171,Prl522111,Pla133171,Prl95130501}. These
can provide a robust way towards universal quantum computation,
because the geometric phases are determined by the global properties of the evolution paths and
possess a
built-in noise-resilience feature against certain types of local
noises \cite{Pra70042316,Pra72020301,Njp14093006,Pra86062322}.
{In particular, nonadiabatic holonomic quantum computation (NHQC) \cite{Pra101022330,Pra95043608,Pra93040305,Prl123100501,Pra101032322,Prl111050404,Prappl14034038}
releases the variations of
parameters from the limitation of the adiabatic condition, 
making the computation fast and robust against local parameter fluctuations over the
cyclic evolution.}
However, due to the
huge physical resource overhead and the difficulties in scaling
up the number of qubits \cite{Prl102070502,Prappl10054051,Prappl13014055,Pra97022335,Pra86032324,arXiv201209034,npjQI11,Nat432602,Sci3321059,Nat51966,Nat506204},
previous work \cite{Prl89097902,Njp14103035,Prl109170501,Pra89042302,Pra95062308,Pra95043608,Pra93040305,Prappl7054022,Pra101022330,Prl123100501,Pra101032322,Prl111050404,Prappl14034038,Nat496482,Prl110190501,Prappl12024024,Nat5147520,Nphoto11309,Prl121110501,Prappl14044043,Prl124230503} showed that is experimentally difficult to implement quantum error
correction protocol \cite{gottesman2010introduction,Pra52R2493,Prsa4522551,Pra86032324} in NHQC.
For this reason, holonomic computation via bosonic codes \cite{Pra561114,Pra64012310} has
attracted much interest recently \cite{Prl116140502,Qst4035007}.
Bosonic codes allow quantum error correction extending only the number
of excitation instead of the number of qubits, while keeping the noise channels fixed \cite{Pra97032346,Njp16045014,Prl119030502,Qst4035007,Fr150,Prl116140502,Prx6031006,Prl124120501,Prl124120501,Nc894,Nc978,Np14705,Nat561368,Np15503,Sci342607,Sci347853,Nc9652,Nat584205}.
For instance, binomial codes \cite{Prx6031006} formed from superposition of Fock states
are protected
against continuous dissipative evolution under loss, gain,
and dephasing errors.

Unfortunately, universal control of a single bosonic
mode is difficult due to its harmonicity.
Although adding direct and indirect nonlinear interactions
can induce weak anharmonicity \cite{Fr150},
it is still difficult to manipulate independently and simultaneously
every needed Fock state. Moreover, weak nonlinear interactions may induce additional noises
into the system and limit the gate fidelities \cite{Fr150}.
This, with additional operations (e.g., feedback \cite{Nat536441,Np15503,Nat584368}
and driven-dissipative controls \cite{Njp16045014,Prl116140502})
and conditions (e.g., oscillators and qubits are never driven simultaneously \cite{Pra92040303,Prl115137002,Prl124120501}), makes it difficult to implement NHQC \cite{Pra101022330,Prl123100501,Pra101032322} with bosonic error-correction codes. 
Note that the first experiment for binomial-code conditional geometric gates
was recently realized \cite{Prl124120501} using 3D superconducting cavities, but it is not
a holonomic computation. 
%The needed nonlinear interactions make
%this protocol slow and reduce the fidelities of the conditional geometric gates.

{The eigenstates of a two-level atom and a cavity field interacting in the ultrastrong coupling (USC) regime are anharmonic dressed atom-light states \cite{Jmp46042311,Prl99173601,Prl107100401,Nr119,Rmp91025005,Epjd59473,Prl98103602,Pra81042311,Prl98103602,Pra81042311,Prl112016401,Prl105263603,
		Prl117043601,Pra96063820,Pra98062327,Np15803,Pra96013849,Pra82022119,Prl116113601,
		Prl119053601,Prl122190403,Pra84043832}. In this manuscript, to overcome the problems mentioned in the previous paragraph, we use these dressed states as intermediate states \cite{Prl110243601,Njp19053010,Pra89033827,Pra94012328,npjQI667,Jpsj88061011} to simultaneously couple different Fock basis and induce population transitions between them. To implement NHQC with binomial codes, we populate Fock states in one step, driving the atom with a composite pulse. The strong anharmonicity in the USC regime allows one to apply strong driving fields \cite{Pra89033827,npjQI667,Pra94012328} in order to shorten the gate time to nanoseconds. These drives are designed by an invariant-based method \cite{Rmp91045001,Aamo62117,Prl116230503,Prl126023602,Jmp101458,Pra86033405,Prl116230503,Pra89033856} and a systematic-error-sensitivity nullification method  \cite{Njp14093040,Pra101032322,Prl111050404}, making our protocol fast and robust against pulse imperfections.
Additionally, the NHQC protocol presented here is scalable for multi-qubit gates
ultrastrongly coupling the atom to a multi-mode cavity. }
%This
%can circumvent the fault-tolerant threshold \cite{Pra52R2493,Prsa4522551,Fr150} when the single-qubit
%gate is extended to multi-qubit gates.}

\begin{figure}
	\centering
	\scalebox{0.086}{\includegraphics{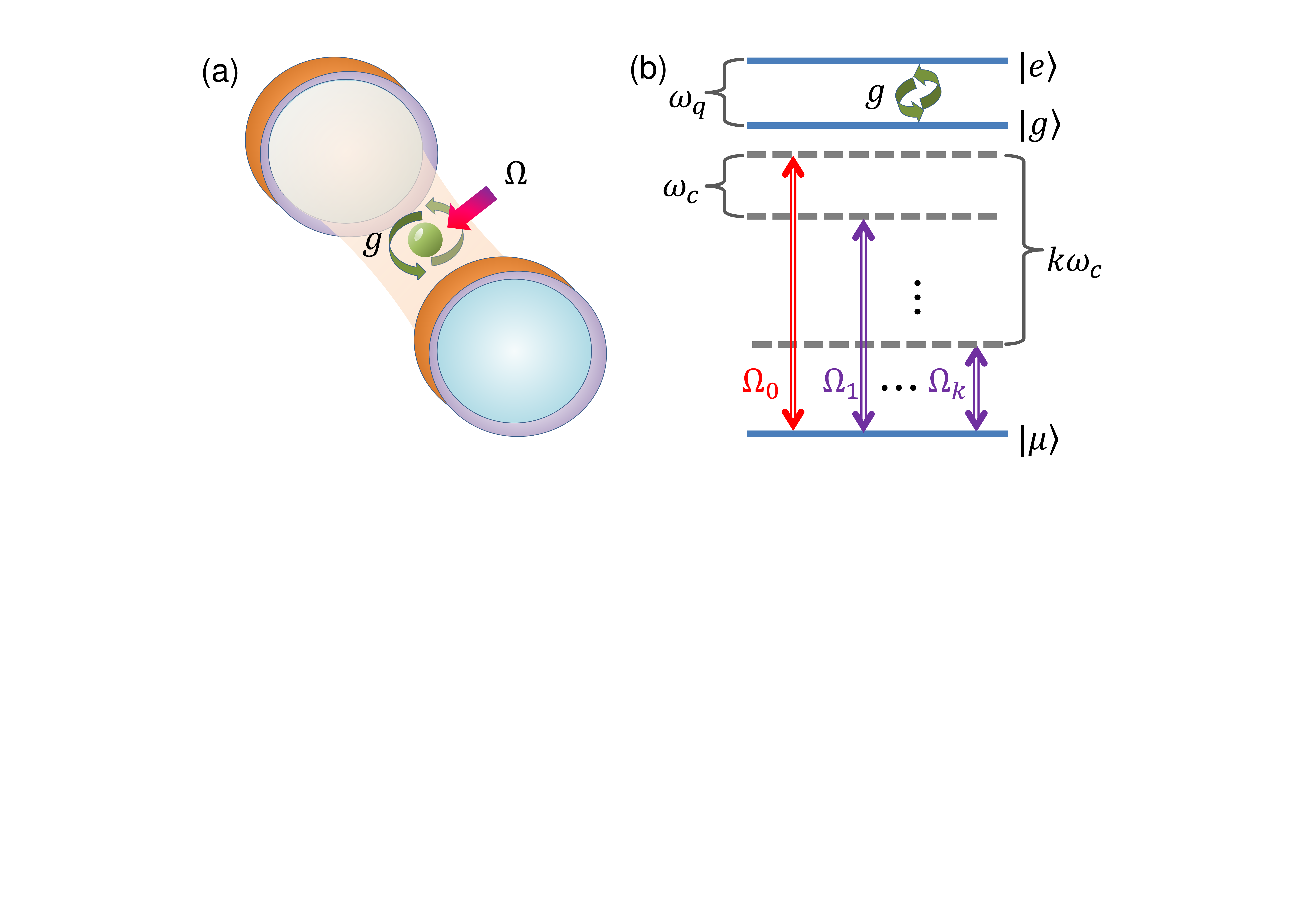}}
	\caption{(a) Schematic illustration of an atom-cavity combined
			system.
			(b) Level diagram of the bare three-level atom. The upper two levels ($|e\rangle,|g\rangle$) of the atom
			are ultrastrongly coupled to the cavity mode with strength $g$. The lower two
			levels ($|g\rangle,|\mu\rangle$) are off-resonantly driven by a composite pulse  $\Omega=\sum_{k}\Omega_{k}\cos{(\omega_{k}t+\phi_{k})}$.			
	}
	\label{figmodel}
\end{figure}

\section{Model and effective Hamiltonian}

Our system consists of
a three-level ($|e\rangle$, $|g\rangle$, $|\mu\rangle$) artificial atom
and a cavity resonator \cite{Epl9924003}. The states $|e\rangle$ and $|g\rangle$
are ultrastrongly coupled to
a cavity mode \cite{Np1344}, with coupling strength $g$ (see Fig.~\ref{figmodel}).
The atom-cavity interaction is described by $H_{0}=H_{R}+\hbar \omega_{\mu}|\mu\rangle\langle\mu|$,
where
\begin{align}\label{eq1-1}
H_{{R}}=\hbar\omega_{{c}}a^{\dag}a+\frac{\hbar\omega_{{q}}}{2}\sigma_{g}^{z}+\hbar g(a+a^{\dag})\sigma_{g}^{x},
\end{align}
is the Rabi Hamiltonian.
Here, $\sigma_{g}^{x}=|e\rangle \langle g|+|g\rangle\langle e|$ and $\sigma_{g}^{z}=|e\rangle\langle e|-|g\rangle\langle g|$
are Pauli matrices,
$a$ ($a^{\dag}$) is the annihilation (creation) operator of the cavity field, $\omega_{\mu}$ is
the frequency of the level $|\mu\rangle$, $\omega_{{c,(q)}}$ is the cavity (qubit) frequency.
In the USC regime ($g/\omega_{c}\gtrsim 0.1$),
%the eigenvectors and eigenvalues of $H_{0}$ are respectively $\left\{|\mathcal{E}_{j}\rangle\right\}=\left\{|\mu\rangle|n\rangle,|\zeta_{m}\rangle\right\}$
%and $\left\{\xi_{j}\right\}=\left\{\omega_{\mu}+n\omega_{c},E_{m}\right\}$ ($j,n=0,1,2,\ldots$ and %$m={0,1,2,}\ldots$).
the eigenstates $|\mathcal{E}_{j}\rangle$ with eigenvalues $\xi_{j}$
of $H_{0}$ can be separated into (i) noninteracting sectors $|\mu\rangle|n\rangle$
with eigenvalues $\omega_{\mu}+n\omega_{{c}}$; and
(ii) dressed atom-cavity states $|\zeta_{m}\rangle$ with eigenvalues $E_{m}$ ($j,n,m=0,1,2,\ldots$).
Here, $|n\rangle$ denote the Fock states of the cavity mode, and
\begin{align}\label{eq2}
  |\zeta_{m}\rangle=\sum_{n}\left(c_{n}^{m}|g\rangle|n\rangle+d_{n\pm 1}^{m}|e\rangle|n\pm 1\rangle\right),
\end{align}
denote the dressed states of $H_{R}$.
The coefficients $c_{n}^{m}=\langle\zeta_{m}|g\rangle|n\rangle$ and $d_{n\pm 1}^{m}=\langle\zeta_{m}|e\rangle|n\pm 1\rangle$
can be obtained numerically.
Note that we impose $d_{-1}^{m}=0$ for Eq.~(\ref{eq2}).

%By introducing a third ancillary level $|\mu\rangle$ at a lower energy $\omega_{\mu}$ [see Fig.~\ref{figmodel}(b)],
%the whole system is diagonalized in the Hilbert space $\left\{|\mathcal{E}_{j}\rangle\right\}\equiv %\left\{|\mu\rangle|n\rangle,|\zeta_{m}\rangle\right\}$ \cite{PM2}.

%the eigenstates $|\mathcal{E}_{j}\rangle$ with corresponding eigenvalues $\xi_{j}$ ($j=0,1,2,\ldots$) of the %whole system are $\left\{|\mathcal{E}_{j}\rangle\right\}\equiv \left\{|\mu\rangle|n\rangle,|\zeta_{m}\rangle\right\}$ \cite{PM2}.

\begin{figure}
	\centering
	\scalebox{0.48}{\includegraphics{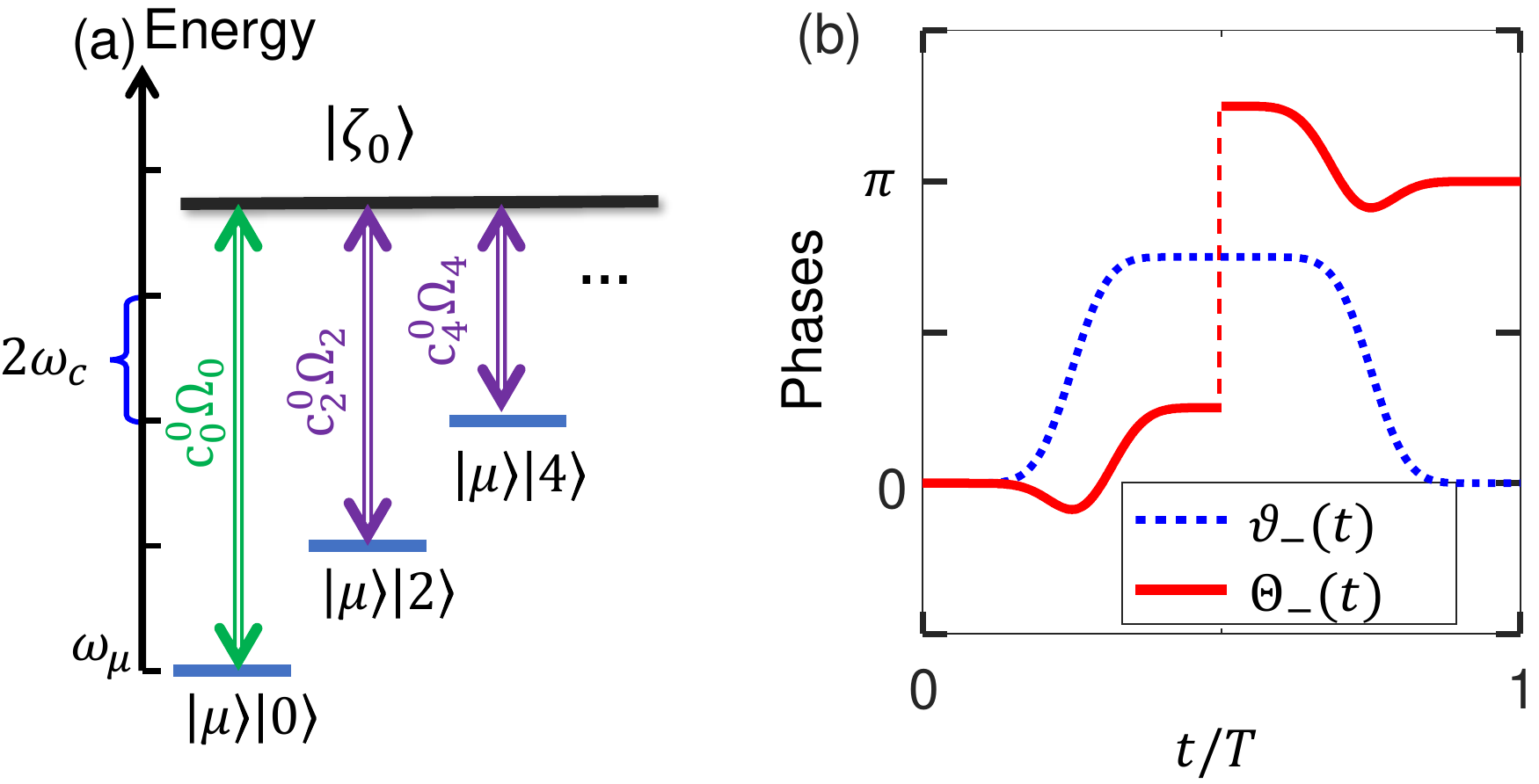}}
	\caption{(a) Illustraton of the effective transitions according to Eq.~(\ref{eq1-2}).
			 (b)  Dynamical and geometric phases acquired by the evolution along $|{\psi}_{-}(t)\rangle$ with parameters
			 given in Eq.~(\ref{eq7a}) and $\Theta_{{s}}=\pi/2$.	
	}
	\label{figefft}
\end{figure}

Oscillations $|\mu\rangle|n\rangle\leftrightarrow|\zeta_{m}\rangle$ can be
induced by driving the atomic transition $|{\mu}\rangle\leftrightarrow|{g}\rangle$ [see Fig.~\ref{figmodel}(b)]
with an additional control Hamiltonian 
\begin{align}
  H_{{D}}(t)=\hbar\Omega(|\mu\rangle\langle g|+|g\rangle\langle\mu|).
\end{align} Here,  
\begin{align}
\Omega=\sum_{k}\Omega_{k}\cos{(\omega_{k}t+\phi_{k})},
\end{align} is a composite pulse \cite{Prl118223604,Sa5eaau5999,Pra89033827,Nat536441} with amplitudes $\Omega_{k}$, frequencies $\omega_{k}$, and phases $\phi_{k}$. We omit the explicit time
dependence of all the parameters (e.g., $\Omega_{k}$ and $\phi_{k}$) regarding the drivings.
{The total Hamiltonian is $H_{\rm{tot}}(t)=H_{0}+H_{{D}}(t)$.}
Choosing 
\begin{align}
\omega_{k}=&~(E_{m}-\omega_{\mu}-k\omega_{{c}})\cr
\Omega_{k}\ll&~\omega_{{c}},g,
\end{align}
and performing a unitary transformation $\exp{\left(-iH_{0}t\right)}$,
we derive an effective Hamiltonian that, under the rotating wave approximation, is (see details in Appendix~\ref{AA})
\begin{align}\label{eq1-2}
  H_{\rm{eff}}(t)=\frac{\hbar}{2}\sum_{k=0}^{k_{\rm{max}}}{c_{k}^{m}\Omega_{k}}e^{i\phi_{k}}|\mu\rangle|k\rangle\langle\zeta_{m}|+{\rm{H.c.}}.
\end{align}
This effective Hamiltonian describes transitions between the Fock states $|k\rangle$ through the dressed intermediate state $|\zeta_{m}\rangle$.
We assume 
\begin{align}
  \omega_{{\mu}}=E_{m}-(k_{\rm{max}}+0.25)\omega_{{c}},
\end{align} 
so
that the dressed state $|\zeta_{m}\rangle$ is the highest level
in the evolution subspace.
In Fig.~\ref{figefft}(a), we illustrate the effective transitions for $m={0}$.
{Note that each Fock state can be freely populated by the drivings $\Omega_{k}$
	when the system is in the USC regime. Instead, in the weak-coupling regime, the qubit driving $H_{D}(t)$ only induces oscillations $|g\rangle|0\rangle
\leftrightarrow|\mu\rangle|0\rangle$ because $c_{n\neq0}^{{m=0}}\simeq0$.
}

\section{Nonadiabatic holonomic quantum computation via binomial codes}

An example of the binomial codes \cite{Prx6031006} for single-qubit gates
protecting against the single-photon loss error is
\begin{align}\label{eq7}
|\tilde{1}\rangle=|2\rangle,\ \ \ \ |\tilde{0}\rangle=(|0\rangle+|4\rangle)/\sqrt{2},
\end{align}
which form a computational subspace $\mathcal{S}_{{c}}=\{|\tilde{0}\rangle,|\tilde{1}\rangle\}$.
With this definition, a photon loss error brings the logical code words to a
subspace with odd photon numbers that is clearly disjoint
from the even-parity subspace of the logical code words \cite{Prx6031006}.
The Knill-Laflamme condition \cite{Pra543824,Pra55900} for this kind of codes reads
$\langle\tilde{\varrho}|a^{\dag}a|\tilde{\varrho}'\rangle=2$ ($\varrho,\varrho'=0,1$).
This means that the probability of a photon jump to
occur is the same for $|\tilde{0}\rangle$ and $|\tilde{1}\rangle$,
implying that the quantum state is not deformed under the
error of a photon loss.
For instance, when encoding quantum
information as 
\begin{align}
  |{\psi}_{0}\rangle=\cos{\chi}|\tilde{0}\rangle+\sin{\chi}|\tilde{1}\rangle,
\end{align}
a photon jump leads to
\begin{align}
  |{\psi}_{1}\rangle=\frac{a|{\psi}_{0}\rangle}{\sqrt{\langle{\psi}_{0}|a^{\dag}a|{\psi}_{0}\rangle}}=\cos\chi|3\rangle+\sin{\chi}|1\rangle,
\end{align} 
which means that the information ($\cos\chi$ and $\sin\chi$) is not
deformed \cite{Prx6031006}.

{To manipulate the codes in Eq.~(\ref{eq7}),
we need a three-frequency composite pulse, i.e., $k=(0,2,4)$ in Eq.~(\ref{eq1-2}).
When $g/\omega_{c}\gtrsim 0.5$, the probability amplitudes $\left(c_{0}^{{2}},c_{2}^{{2}},c_{4}^{{2}}\right)$ of the Fock states $\left(|0\rangle,|2\rangle,|4\rangle\right)$
in the third dressed state $|\zeta_{{2}}\rangle$ are greater than in the other dressed states (see more details in Appendix~\ref{AA}).
For this reason, we choose $m={2}$ in Eq.~(\ref{eq1-2}).
Assuming $c_{0}^{{2}}\Omega_{0}=c_{4}^{{2}}\Omega_{4}$ and $\phi_{0}=\phi_{4}$, $H_{\rm{eff}}(t)$ becomes an effective
$\Lambda$-type system with two ground states $\{|\tilde{0}\rangle,|\tilde{1}\rangle\}$ and an excited state $|\zeta_{2}\rangle\equiv|\zeta_{{m=2}}\rangle$.
The NHQC in a $\Lambda$-type system has been well studied \cite{Njp14103035,Prl109170501,Pra72020301}.} For clarity, we
define an effective driving amplitude 
\begin{align}
  {\Xi}=\sqrt{\sum_{k}{\left(c_{k}^{{2}}\Omega_{k}\right)^2}},
\end{align}
and
a time-independent parameter 
\begin{align}
  \theta=\frac{1}{2}\arctan\left(\frac{\sqrt{2}c_{0}^{{2}}\Omega_{0}}{c_{2}^{{2}}\Omega_{2}}\right),
\end{align} 
to rewrite $H_{\rm{eff}}(t)$ to be
\begin{eqnarray}\label{eq9}
{H}_{\rm{eff}}(t)=\frac{\hbar}{2}{\Xi}\exp{\left({i{\phi}_{2}}\right)}|b\rangle\langle\zeta_{{2}}|+{\rm{H.c.}},
\end{eqnarray}
where $\phi=\phi_{2}-\phi_{0}$ and 
\begin{align}
  |b\rangle=e^{-i\phi}\sin({\theta}/{2})|\tilde{0}\rangle|\mu\rangle+\cos({\theta}/{2})|\tilde{1}\rangle|\mu\rangle,
\end{align}
%{According to the invariant-based STA approaches \cite{Pra83062116,Pra89033856,SM,Pra86033405}
%for $\Lambda$-type transitions, we can define a pump pulse $\Omega_{p}(\beta,\varphi)\equiv(\dot{\beta} \cot %\varphi \sin \beta+\dot{\varphi} \cos \beta)$ and a Stokes pulse
%$\Omega_{s}(\beta,\varphi)\equiv(\dot{\beta} \cot \varphi \cos
%\beta-\dot{\varphi} \sin \beta)$, where $\beta$ and $\varphi$ are time-dependent auxiliary
%parameters to be determined.}

Initially, quantum information is stored in the logical qubit states of
the subspace $\mathcal{S}_{{c}}$ (the atom is in $|\mu\rangle$).
According to the invariant-based approaches 
for $\Lambda$-type transitions \cite{Pra83062116,Pra89033856,Pra86033405}, when
\begin{align}
  {\Xi}\sin{\phi_{2}}&=\Omega_{{p}}(\beta,\varphi)\equiv(\dot{\beta} \cot \varphi \sin \beta+\dot{\varphi} \cos \beta), \cr
  {\Xi}\cos{\phi_{2}}&=\Omega_{{s}}(\beta,\varphi)\equiv(\dot{\beta} \cot \varphi \cos \beta-\dot{\varphi} \sin \beta),
\end{align}
the Hamiltonian in Eq.~(\ref{eq9}) can drive the system
to evolve exactly along one of the two user-defined path (see details in Appendix~\ref{AB})
\begin{align}
 |\psi_{+}(t)\rangle=&\sin({\varphi}/2)|\mu\rangle|b\rangle+i\exp({i{\beta}})\cos({\varphi}/2)|\zeta_{{2}}\rangle,\cr
 |\psi_{-}(t)\rangle=&i\exp({i{\beta}})\cos({\varphi}/2)|b\rangle+\sin({\varphi}/2)|\zeta_{{2}}\rangle,
\end{align}
which are two eigenstates of a dynamical invariant $I(t)$
obeying $\hbar{\partial_{t}}I(t)=i[H_{\rm{eff}}(t),I(t)]$. 
For instance, when $\varphi(0)=0$, the evolution is along $|\psi_{-}(t)\rangle$, which acquires
a dynamical phase
%\begin{align}\label{eq1-7}
\begin{align}
  \vartheta_{-}(t)=-\frac{1}{\hbar}\int_{0}^{t}\langle{\psi_{-}(t')}|H_{\rm{eff}}(t')|{\psi_{-}(t')}\rangle dt',
\end{align} 
and a geometric phase
\begin{align}
  \Theta_{-}(t)=\int_{0}^{t}\langle{\psi_{-}(t')}|{i\partial_{t'}}|{\psi_{-}(t')}\rangle dt'.
\end{align}

{For a cyclic evolution, the time-dependent auxiliary parameters ${\beta}$ and ${\varphi}$ 
need to satisfy  $\beta(0)\neq \beta(t_{f})$ and
${\varphi}(0)={\varphi}(t_{{f}})=0$.
We can choose  
\begin{align}\label{eq7a}
{\varphi}=&\pi\sin^{2}(\pi t/T),\cr\cr
{\beta}=&\frac{2}{3}\left\{\begin{array}{ll}
2\sin^{3}{\varphi},&             \ \ \ \     t\in[0,t_{{f}}/2] \\
2\sin^{3}{\varphi}-3\Theta_{{s}}.&\ \ \ \ t\in[t_{{f}}/2,t_{{f}}]
\end{array}
\right.
\end{align}
Thus, the final phases are $\vartheta_{-}(t_{f})=0$ and $\Theta_{-}(t_{f})=2\Theta_{s}$ [see Fig.~\ref{figefft}(b) and Appendix~\ref{AB}], resulting in a geometric evolution. 
In the computational subspace $\mathcal{S}_{{c}}$, the evolution operator is (omitting a global phase $\Theta_{s}$)
\begin{eqnarray*}\label{eq10}
	U_{T}=\left(\begin{array}{cc}
		\cos{\Theta_{s}}+i\sin{\Theta_{s}}\cos{\theta} &
		i\sin{\Theta_{s}}\sin{\theta}e^{i\phi} \\
		i\sin{\Theta_{s}}\sin{\theta}e^{-i\phi}  &
		\cos{\Theta_{s}}-i\sin{\Theta_{s}}\cos{\theta}
	\end{array}
	\right).
\end{eqnarray*}
This is a universal single-qubit gate.

In this case, the gate time is $T\sim 18 /c_{2}^{k}\Omega_{k}^{\rm{peak}}$. %satisfying $T\gg 18/c_{2}^{k}\omega_{c}$, due to $\omega_{{c}},g\gg\Omega_{k}$.
In the USC regime we can assume $c_{k}^{m}\gtrsim 0.1$ and $\omega_{c}/2\pi\sim 5~{\rm{GHz}}$ \cite{Nr119,Rmp91025005}, resulting in $T\gg 5~{\rm{ns}}$, i.e., the gate time can be tens of nanoseconds.
Choosing $T=35~{\rm{ns}}$, $g\simeq 0.8\omega_{c}$ and $\omega_{c}/2\pi=6.25~{\rm{GHz}}$ \cite{Np1344}, the pulses $\Omega_{0,(2,4)}$ are shown in Fig.~\ref{figPF}(a).
Note there that the peak values of the pulses are $\Omega_{k}^{\rm{peak}}/2\pi\sim 200~{\rm{MHz}}$. These
satisfy the condition $\omega_{{c}},g\gg\Omega_{k}$.

}

{\section{Robustness against control imperfections and decoherence}

It has been experimentally verified \cite{Prappl14054062} that
the pulses chosen based on $\beta$ and $\varphi$ in Eq.~(\ref{eq7a}) can counteract the systematic errors induced by imperfections of the control fields $\Omega_{k}$,
making the computation insensitive to such errors \cite{Njp14093040,Pra101032322,Prl111050404}.
In the presence of such imperfections with error
parameter $\delta_{{i}}$, the driving amplitudes become
$\Omega_{k}^{{i}}=(1+\delta_{{i}})\Omega_{k}$. Accordingly,
the effective Hamiltonian ${H}_{\rm{eff}}(t)$
should be corrected as ${H}_{\rm{eff}}^{{i}}(t)=(1+\delta_{{i}}){H}_{\rm{eff}}(t)$. By using time-dependent perturbation theory up to $\mathcal{O}(\delta_{{i}})$, the
evolution state of the system is approximatively
\begin{eqnarray*}
	|{\psi}_{-}^{{i}}(t)\rangle\approx 
	|{\psi}_{-}(t)\rangle-\frac{i\delta_{{i}}}{\hbar}\int_{0}^{t_{{f}}}U(t_{{f}},t){H}_{\rm{eff}}^{{i}}(t)|{\psi}_{-}(t)\rangle dt
\end{eqnarray*}
where $U(t_{{f}},t)$ is the unperturbed time evolution operator.
%Here, for simplicity, we have assumed that the evolution is along $|{\psi}_{-}(t)\rangle$
%by designing ${\varphi}(0)={\varphi}(t_{{f}})=0$.

\begin{figure}
	\centering
	\scalebox{0.48}{\includegraphics{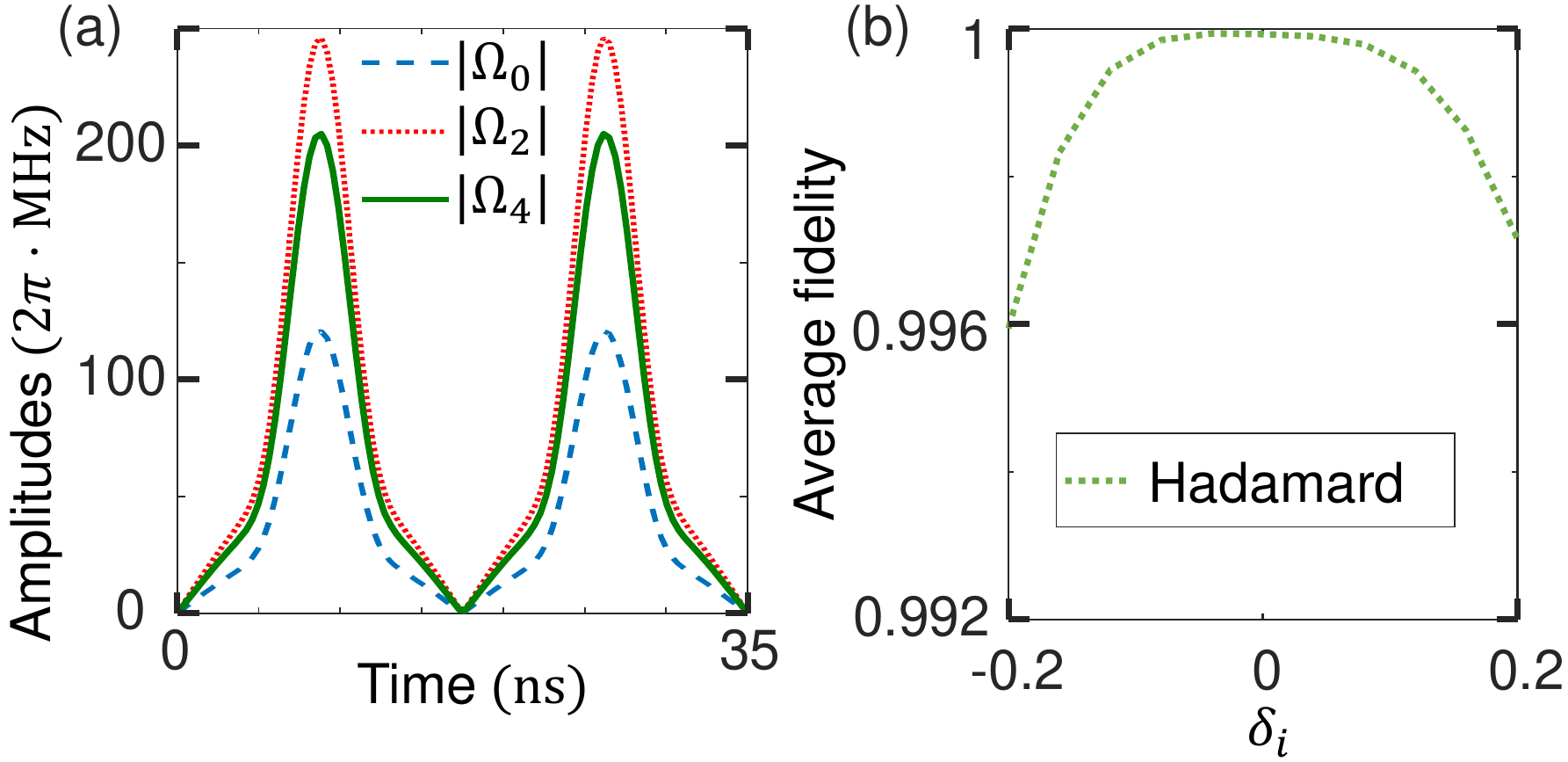}}
	\caption{
		{For $(\Theta_{{s}},\theta,\phi)=(\pi/2,\pi/4,0)$, (i.e., the Hadamard gate)}: (a) Finite-duration drives $\Omega_{0}=\left(\Xi/\sqrt{2}c_{0}^{{2}}\right)\sin(\theta/2)$,
		$\Omega_{2}=\left(\Xi /c_{2}^{{2}}\right)\cos(\theta/2)$, and $\Omega_{4}=\left(\Xi/\sqrt{2}c_{4}^{{2}}\right)\sin(\theta/2)$ when $T=35~{\rm{ns}}$.
		(b) Average fidelity $\bar{F}$ of the Hadamard versus the error coefficient $\delta_{{i}}$. The intermediate state is $|\zeta_{{2}}\rangle$ and the gate time is $T=150~{\rm{ns}}$. Other parameters are $g=0.8\omega_{{c}}$ and $\omega_{{q}}=\omega_{{c}}=2\pi\times6.25~{\rm{GHz}}$ \cite{Np1344}.
	}
	\label{figPF}
\end{figure}

We assume that the protocol works perfectly when $\delta_{{i}}=0$,
resulting in
\begin{eqnarray*}\label{eqS23}
	P_{\rm{out}}\approx 1-\frac{\delta_{{i}}^2}{\hbar{^2}}\left|\int_{0}^{t_{{f}}} e^{2i\mathcal{R}_{-}(t)}\langle{\psi}_{+}(t)|{H}_{\rm{eff}}(t)|{\psi}_{-}(t)\rangle dt\right|^{2},
\end{eqnarray*}
where $P_{\rm{out}}$ is the population of the output state
after the gate operation and 
\begin{align}
\mathcal{R}_{-}(t)=&\frac{1}{\hbar}\int_{0}^{t}\langle{\psi_{-}(t')}|\left[i{\hbar}{\partial_{t'}}-H_{\rm{eff}}(t')\right]|{\psi_{-}(t')}\rangle dt',
\end{align}
is the Lewis-Riesenfeld phase \cite{Jmp101458}. 
Then, the systematic error sensitivity can be defined as \cite{Njp14093040}
\begin{align}\label{eqS22}
q_{{i}}:=&-\left.\frac{1}{2}\frac{\partial^2 P_{\rm{out}}}{\partial \delta_{{i}}^2}\right|_{\delta_{{i}}=0} \cr
=&\left|\int_{0}^{t_{{f}}}e^{i{\beta}+2i\mathcal{R}_{-}(t)}\dot{{\varphi}}\sin^{2}{\varphi} dt\right|^{2}.
\end{align}
Substituting $\varphi$ and $\beta$ [see Eq.~(\ref{eq7a})] into Eq.~(\ref{eqS22}),
we obtain $q_{{i}}\simeq 0$ \cite{Njp14093040,Pra101032322,Prl111050404}, which means that the holonomic gates
are insensitive to the systematic errors induced by the pulse imperfections. }

\begin{figure}
	\centering
	\scalebox{0.42}{\includegraphics{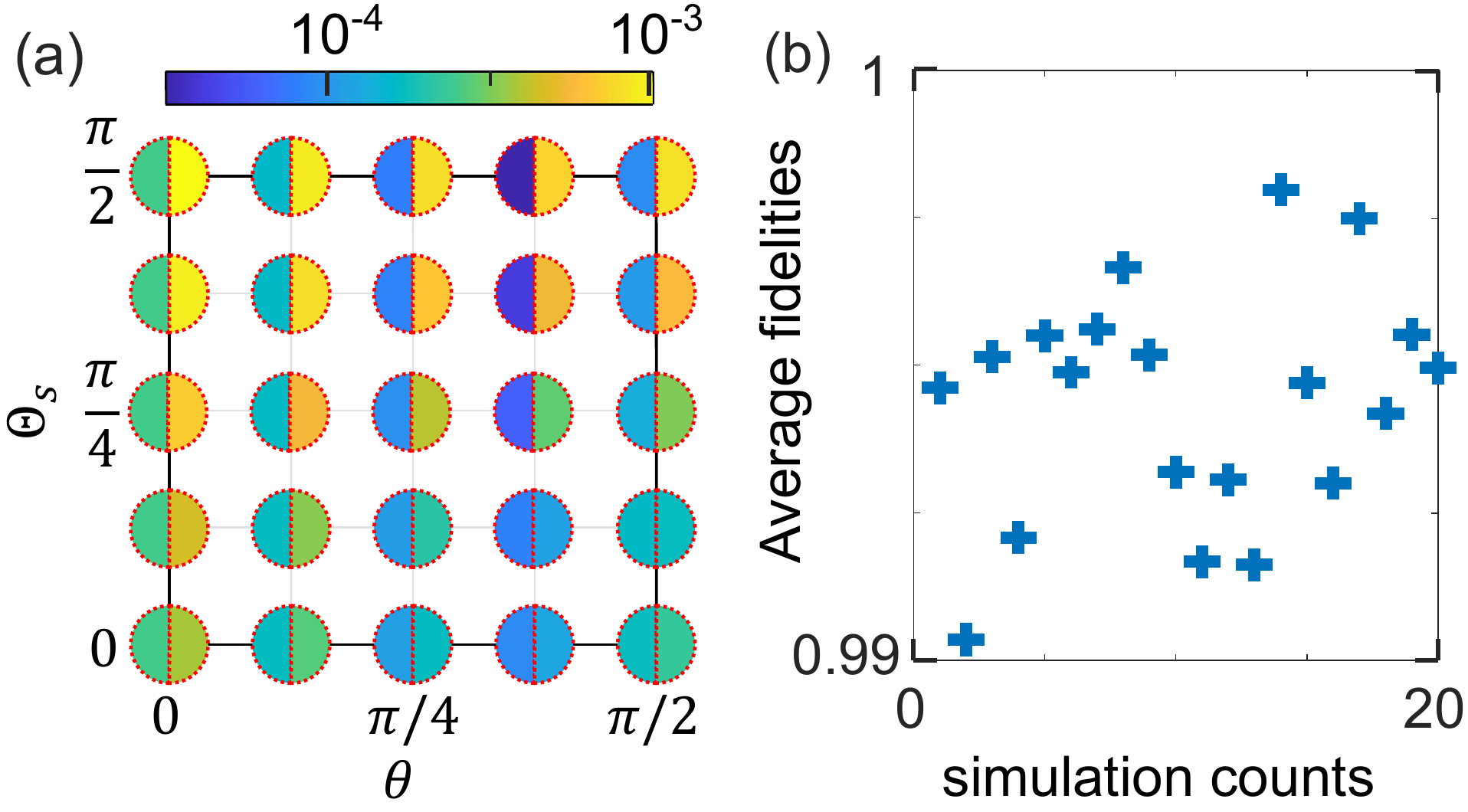}}
	\caption{(a) Average infidelities $(1-\bar{F})$ of arbitrary single-qubit gates when $\phi=0$ and $T=150~{\rm{ns}}$. Each circle denotes a single-qubit gate, e.g.,
		$(\Theta_{{s}},\theta)=(\pi/2,\pi/4)$ corresponds to the
		Hadamard gate. In each circle, the left (right) side denotes the infidelity
		in the absence (presence) of pulse imperfections. When considering pulse imperfections, the error coefficient is assumed to be $\delta_{{i}}=0.1$.
		(b) Average fidelities $\bar{F}$ of the Hadamard gate versus simulation counts with an additive white Gaussian noise (AWGN). 
		Each cross denotes the average gate fidelity when the signal-to-noise ratio is $r=15$.
		Other parameters are the same as those in Fig.~\ref{figPF}.
	}
	\label{fig4}
\end{figure}

The average fidelity of a gate
over all possible initial states can be defined by \cite{Pra70012315,Pla36747}
\begin{align}\label{e15}
\bar{F}=\left[\mathrm{Tr}(MM^\dag)
+|\mathrm{Tr}(M)|^2\right]/(D^2+D),
\end{align}
with $M=\mathcal{P}_{{c}}U^\dag_{T}U\mathcal{P}_{{c}}$. Here, $\mathcal{P}_{{c}}$ ($D$) is the projector
(dimension) of the subspace $\mathcal{S}_{{c}}$.
The evolution operator $U$, describing the actual dynamical evolution,
is calculated with the total Hamiltonian $H_{\rm{tot}}(t)=H_{0}+H_{D}(t)$.
{Using the above definition, in Fig.~\ref{figPF}(b), we show the average fidelity $\bar{F}$ of the Hadamard gate
versus the error coefficient $\delta_{{i}}$. Note that,
when $\delta_{{i}}\in[-0.1,0.1]$, the average fidelity is nearly $99.9\%$, indicating that
our protocol is insensitive to the systematic error caused by pulse imperfections. }

The average infidelities ($1-\bar{F}$) of arbitrary single-qubit gates are shown
in Fig.~\ref{fig4}(a). The left (right) side
of each circle denotes the average infidelity
in the absence (presence) of pulse imperfections.
When considering pulse imperfections with an error coefficient $\delta_{{i}}=0.1$,
the infidelities only slightly increase from $\sim10^{-4}$ to $\sim 10^{-3}$.
For instance, in the case of the Hadamard gate, pulse imperfections with an error coefficient $\delta_{i}=0.1$
only increase the infidelity from $<10^{-4}$ to $\sim10^{-3}$.
This indicates that
the generated gates are mostly insensitive to systematic errors.

{Generally, a geometric gate can be robust against 
noise caused by amplitude fluctuations. 
Without loss of generality, we use additive white
Gaussian noise to investigate the influence of such noise.
In this case, the driving amplitudes $\Omega_{k}$ should be
corrected to be $\Omega_{k}^{s}={\rm{AWGN}}(\Omega_{k},r)$.
Here, ${\rm{AWGN}}(\Omega_{k},r)$ is a
function that generates the additive white Gaussian noise (AWGN)
to the original signal 
$\Omega_{k}$ with a signal-to-noise ratio $r$.
Because the additive white Gaussian noise is generated
randomly in each single simulation, we perform the
numerical simulation 20 times to estimate its average
influence [see Fig.~\ref{fig4}(b) with an illustraton of the Hadamard gate]. 
As shown in Fig.~\ref{fig4}(b), when considering 
relatively strong noises with $r=15$, the gate fidelities
can still be higher than $99\%$. This
indicates that our protocol is mostly insensitive to noise caused by amplitude fluctuations. 
}

In the USC regime,
relaxation and dephasing are
studied in the basis $|\mathcal{E}_{j}\rangle$,
which diagonalizes the Hamiltonian $H_{0}$.
The master equation in the
Born-Markov approximation, valid for generic hybrid-quantum systems, is \cite{Pra84043832,Prl110243601,Njp19053010,Pra89033827,Pra97033823}
\begin{eqnarray}\label{eq1-6}
\hbar \dot{\rho}(t)&=&i[\rho(t),H_{\rm{tot}}(t)]+
\sum_{\nu=0}^{3}\mathcal{D}\left[\sum_{j}\sqrt{\Lambda_{\nu}^{jj}}|\mathcal{E}_{j}\rangle\langle\mathcal{E}_{j}|\right]\rho(t)
\cr&&+\sum_{\nu'=0}^{5}\sum_{j>j',j'}\Gamma_{\nu'}^{jj'}\mathcal{D}[|\mathcal{E}_{j'}\rangle\langle\mathcal{E}_{j}|]\rho(t),
%+\kappa^{\phi}\mathcal{D}{[X^{-}X^{+}]\rho(t)}
\end{eqnarray}
where $\mathcal{D}[\mathcal{O}]\rho(t)=\mathcal{O}\rho(t)\mathcal{O}^{\dag}-[\rho(t)
\mathcal{O}^{\dag}\mathcal{O}+\mathcal{O}^{\dag}\mathcal{O}\rho(t)]/2$
is the Lindblad superoperator.
For simplicity, the dephasing and relaxation parameters have been written in a compact form:
\begin{align}
\Lambda_{0}^{jj}&=\kappa^{\phi}|\langle\mathcal{E}_{j}|a^{\dag}a|\mathcal{E}_{j}\rangle|^{2}, \cr
\Lambda_{1}^{jj}&=\kappa|\langle\mathcal{E}_{j}|a^{\dag}+a|\mathcal{E}_{j}\rangle|^{2}, \cr
\Lambda_{2,(3)}^{jj}&=\gamma_{g,(\mu)}^{\phi}|\langle\mathcal{E}_{j}|\sigma_{g,(\mu)}^{z}|\mathcal{E}_{j}\rangle|^{2},\cr
\Gamma_{{0}}^{jj'}&=\kappa^{\phi}|\langle\mathcal{E}_{j'}|a^{\dag}a|\mathcal{E}_{j}\rangle|^{2}, \cr
\Gamma_{{1}}^{jj'}&=\kappa|\langle\mathcal{E}_{j'}|a^{\dag}+a|\mathcal{E}_{j}\rangle|^{2},\cr
\Gamma_{2,(3)}^{jj'}&=\gamma_{g,(\mu)}|\langle\mathcal{E}_{j'}|\sigma_{g,(\mu)}^{x}|\mathcal{E}_{j}\rangle|^{2},\cr
\Gamma_{4,(5)}^{jj'}&=\gamma_{g,(\mu)}^{\phi}|\langle\mathcal{E}_{j'}|\sigma_{g,(\mu)}^{z}|\mathcal{E}_{j}\rangle|^{2}.
\end{align}
Here, $\sigma_{\mu}^{x}=|\mu\rangle\langle g|+|g\rangle\langle\mu|$, $\kappa$ ($\kappa^{\phi}$) is the cavity decay (dephasing) rate, $\gamma_{g,(\mu)}$ is the spontaneous emission rate of the transition
$|e\rangle\rightarrow|g\rangle$ ($|g\rangle\rightarrow|\mu\rangle$), and 
$\gamma^{\phi}_{g,(\mu)}$ is the atomic 
dephasing rate corresponding to $\sigma_{g,(\mu)}^{z}$ ($\sigma_{\mu}^{z}=|g\rangle\langle g|-|\mu\rangle\langle \mu|$).

%Here, $\kappa$ ($\kappa^{\phi}$) is the cavity decay (dephasing) rate, %$\gamma_{g,(\mu)}$ is the spontaneous emission rate of the transition
%$|e\rangle\rightarrow|g\rangle$ ($|g\rangle\rightarrow|\mu\rangle$), and
%$\gamma^{\phi}_{g,(\mu)}$ is the atomic
%dephasing rate corresponding to $\sigma_{g,(\mu)}^{z}$ %($\sigma_{\mu}^{z}=|g\rangle\langle g|-|\mu\rangle\langle \mu|$).

\begin{figure}
	\centering
	\scalebox{0.45}{\includegraphics{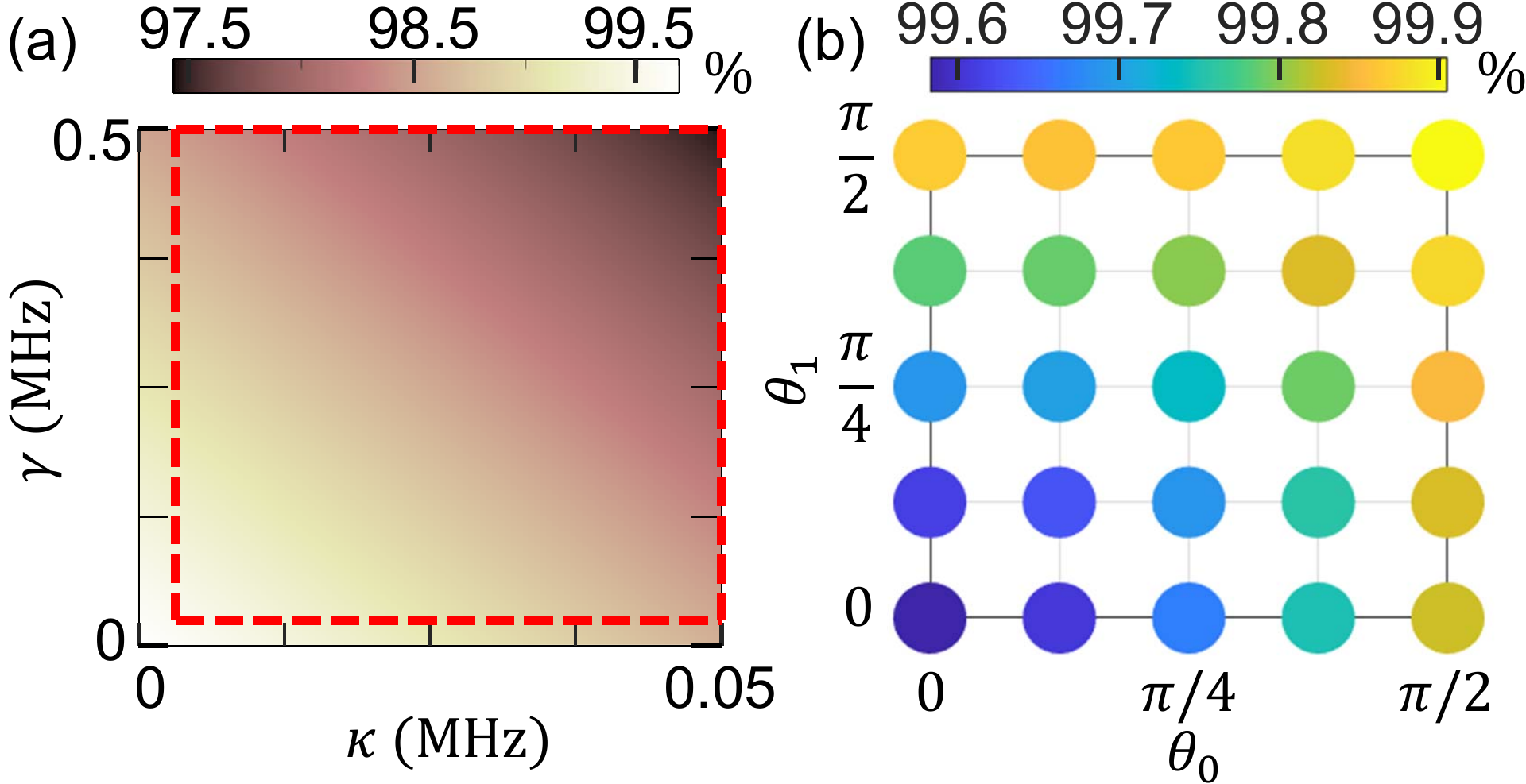}}
	\caption{
		(a) Fidelity $F_{\rm{out}}$ of the output state of the Hadamard gate versus $\kappa$ and $\gamma$.
		The decay and dephasing rates in the red-dashed box have been achieved in experiments of superconducting circuits \cite{Fr150}.
		Other parameters are the same for Fig.~\ref{figPF}.
		(b) Average fidelities of arbitrary two-qubit gates in the presence of pulse imperfections with an error coefficient $\delta_{{i}}=0.1$. We assume 
		$\Theta_{{s}}=\pi/2$ and $\phi=\pi$ for simplicity. Each circle denotes a two-qubit gate, e.g., 
		$(\theta_{0},\theta_{1})=(0,\pi/2)$ corresponds to the controlled-Not gate.
		The intermediate state is $|\zeta'_{0}\rangle$; Parameters are $T=750~{\rm{ns}}$, $g_{a}=g_{b}=1.3\omega_{{c}}$, and $\omega_{b}=0.9\omega_{a}$, $\omega_{a}=\omega_{{q}}=2\pi\times6.25~{\rm{GHz}}$.
	}
	\label{fig5}
\end{figure}

To check the robustness of the geometric gates against decoherence,
we assume the input state as
$|{\psi}_{\rm{in}}\rangle=|\tilde{0}\rangle$, corresponding to
an output state $|{\psi}_{\rm{out}}\rangle=U_{T}|{\psi}_{\rm{in}}\rangle$.
Using $(\Theta_{{s}},\theta,\phi)=(\pi/2,\pi/4,0)$ (Hadamard gate), in Fig.~\ref{fig5}(a) we show the
fidelity
$F_{\rm{out}}=\langle\psi_{\rm{out}}|\rho(t_{{f}})|\psi_{\rm{out}}\rangle$ versus $\gamma$ and $\kappa$
in the presence of pulse imperfections when $\delta_{{i}}=0.1$.
In this figure we notice that the dissipation and dephasing of the atom
affect the evolution much weaker than those of the cavity.
For experimentally realistic parameters of superconducting circuit experiments \cite{Prl124120501}, $\left(\kappa,\kappa^{\phi},\gamma_{g,(\mu)},\gamma_{g,(\mu)}^{\phi}\right)\simeq2\pi\times\left(0.33,0.3,8,8\right)~{\rm{kHz}}$,
the fidelity of the output state is $F_{\rm{out}}\simeq99.56\%$,
indicating that our protocol is robust against
decoherence. 

{\section{Multi-qubit gates}

Our protocol can be extended to implement multi-qubit holonomic gates, such as two-qubit gates.
We consider that the $\Xi$-type atom ultrastrongly couples to a bimodal cavity (frequencies $\omega_{a}$ and $\omega_{b}$). The system Hamiltonian is described by 
\begin{align}
H'_{\rm{tot}}=&H'_{{R}}+\hbar \omega_{\mu}|\mu\rangle\langle\mu|+H_{{D}}(t), \cr\cr
H'_{{R}}=&\hbar\omega_{{a}}a^{\dag}a+\hbar\omega_{{b}}b^{\dag}b+\frac{\hbar\omega_{{q}}}{2}\sigma^{z}_{g}\cr
&+\hbar[g_{a}(a+a^{\dag})+g_{b}(b+b^{\dag})]\sigma^{x}_{g}.
\end{align} 
The eigenstates of $H'_{{R}}$ corresponding to the eigenvalues $E'_{m}$ can be described by
\begin{align*}
|\zeta'_{m}\rangle=\sum_{n_{a},n_{b},m}c^{m}_{n_{a},n_{b}}|g\rangle|n_{a}\rangle_{a}|n_{b}\rangle_{b}
+d^{m}_{n_{a},n_{b}}|e\rangle|n_{a}\rangle_{a}|n_{b}\rangle_{b},
\end{align*}
where $|n_{a}\rangle_{b}$ and $|n_{b}\rangle_{b}$ denote the
Fock states of the two cavity modes, respectively.

{\renewcommand\arraystretch{4.2}
	\begin{table}
		\centering{
			\caption{Implementation examples of two-qubit gates\\ }
			\label{tab2}
			\begin{tabular}{p{1.3cm}<{\centering}|p{4cm}<{\centering}|p{3cm}<{\centering}}
				\hline
				\hline
				gate & matrix & parameters $(\Theta_{{s}},\theta_{0},\theta_{1},\theta_{2},\phi)$ \\
				\hline
				CNot    & \renewcommand{\arraystretch}{1}$\left(\begin{array}{cccc}\setlength{\arraycolsep}{0.1pt}
				0 & 1 & 0 & 0\\
				1 & 0 & 0 & 0\\
				0 & 0 & 1 & 0\\
				0 & 0 & 0 & 1
				\end{array}
				\right)
				$ &       $({\pi}/{2},0,{\pi}/{2},{\pi}/{2},\pi)$    \\
				\hline
				SWAP    & \renewcommand{\arraystretch}{1}$\left(\begin{array}{cccc}
				1 & 0 & 0 & 0 \cr
				0 & 0 & 1 & 0\\
				0 & 1 & 0 & 0\\
				0 & 0 & 0 & 1
				\end{array}
				\right)
				$ &       $({\pi}/{2},-{\pi}/{2},0,\pi,\pi)$    \\
				\hline
				$\sqrt{\rm{SWAP}}$    & {\renewcommand{\arraystretch}{1.2}$\left(\begin{array}{cccc}
					1 & 0 & 0 & 0\\
					0 & \frac{1}{2}(1+i) & \frac{1}{2}(1-i) & 0\\
					0 & \frac{1}{2}(1-i) & \frac{1}{2}(1+i) & 0\\
					0 & 0 & 0 & 1
					\end{array}
					\right)
					$} &       $({\pi}/{4},-{\pi}/{2},\pi,0,\pi)$    \\
				\hline\hline
		\end{tabular}}
	\end{table}

Then, we assume that the driving field is
\begin{align}
\Omega=\Omega_{k_{a},k_{b}}\cos(\omega_{k_{a},k_{b}}+\phi_{k_{a},k_{b}}).
\end{align}
When choosing the frequencies that $\omega_{a}/\omega_{b}\neq 0, 1, 2\ldots$,
and 
\begin{align}
\omega_{k_{a},k_{b}}=E'_{m}-\omega_{\mu}-k_{a}\omega_{a}-k_{b}\omega_{b},
\end{align}
the
effective Hamiltonian is approximatively
\begin{align*}\label{eqS25}
H'_{\rm{eff}}(t)=&\frac{\hbar}{2}\sum_{k_{a},k_{b}}{c^{m}_{k_{a},k_{b}}\Omega_{k_{a},k_{b}}}\exp{\left(i\phi_{k_{a},k_{b}}\right)}|\mu\rangle|k_{a}\rangle_{a}|k_{b}\rangle_{b}\langle\zeta'_{m}|\cr
&+{\rm{H.c.}}.
\end{align*}
For simplicity, we assume that the intermediate state is the dressed state $|\zeta'_{{0}}\rangle$, 
the driving amplitudes become
\begin{align*}
&c^{{0}}_{0,0}\Omega_{0,0}=c^{{0}}_{0,4}\Omega_{0}^{4}=c^{{0}}_{4,0}\Omega_{4,0}=c^{{0}}_{4,4}\Omega_{4,4}=\Xi_{\tilde{0}\tilde{0}}(t)/2,\cr
&c^{{0}}_{0,2}\Omega_{0,2}=c^{{0}}_{4,2}\Omega_{4,2}=\Xi_{\tilde{0}\tilde{1}}(t)/\sqrt{2},\cr
&c^{{0}}_{2,0}\Omega_{2,0}=c^{{0}}_{2,4}\Omega_{2,4}=\Xi_{\tilde{1}\tilde{0}}(t)/\sqrt{2},\cr
&c^{{0}}_{2,2}\Omega_{2,2}=\Xi_{\tilde{1}\tilde{1}}(t),
\end{align*}
and the phases are
\begin{align}
&\phi_{0,0}= \phi_{0,4} =\phi_{4,0} =\phi_{4,4}=\phi_{\tilde{0}\tilde{0}},\cr
&\phi_{0,2}=\phi_{4,2}=\phi_{\tilde{0}\tilde{0}}+\phi,\cr
&\phi_{2,0}=\phi_{2,4}=\phi_{\tilde{0}\tilde{0}}+\phi,\cr
&\phi_{2,2}=\phi_{\tilde{0}\tilde{0}}+\phi.
\end{align}
Here, the auxiliary parameter $\phi_{\tilde{0}\tilde{0}}$ is time-dependent and the auxiliary parameter $\phi$ is time-independent.

}

The effective Hamiltonian in Eq.~(\ref{eqS25}) becomes
\begin{align}
\tilde{H}'_{\rm{eff}}(t)=%&\frac{e^{i\phi_{\tilde{0}\tilde{0}}}}{2}|\mu\rangle\left[\Xi_{\tilde{0}\tilde{0}}(t)|\tilde{0}\rangle_{a}|\tilde{0}\rangle_{b}+\Xi_{\tilde{0}\tilde{1}}(t)e^{i\phi}|\tilde{0}\rangle_{a}|\tilde{1}\rangle_{b}+\Xi_{\tilde{1}\tilde{0}}(t)e^{i\phi}|\tilde{1}\rangle_{a}|\tilde{0}\rangle_{b}+\Xi_{\tilde{1}\tilde{1}}(t)e^{i\phi}|\tilde{1}\rangle_{a}|\tilde{1}\rangle_{b}\right]\langle\zeta'_{0}|+{\rm{H.c.}},\cr
\frac{\hbar}{2}{{\Xi'}_{0}(t)\exp{\left[i\phi_{\tilde{0}\tilde{0}}\right]}}|\mu\rangle|{b'}\rangle\langle\zeta'_{{0}}|+{\rm{H.c.}},
\end{align}
with the binomial codes
\begin{align}
|\tilde{0}\rangle_{a}=&\frac{1}{\sqrt{2}}(|0\rangle_{a}+|4\rangle_{a}),\ \ \
|\tilde{1}\rangle_{a}=|2\rangle_{a},\cr
|\tilde{0}\rangle_{b}=&\frac{1}{\sqrt{2}}(|0\rangle_{b}+|4\rangle_{b}),\ \ \ \
|\tilde{1}\rangle_{b}=|2\rangle_{b}.
\end{align}
Here, the bright state $|b'\rangle$ can be defined as 

\begin{align*}
|b'\rangle=&
e^{-i\phi}\cos{\frac{\theta_{0}}{2}}\cos{\frac{\theta_{1}}{2}}|\tilde{0}\rangle_{a}|\tilde{0}\rangle_{b}
+\cos{\frac{\theta_{0}}{2}}\sin{\frac{\theta_{1}}{2}}|\tilde{0}\rangle_{a}|\tilde{1}\rangle_{b}\cr
&+\sin{\frac{\theta_{0}}{2}}\cos{\frac{\theta_{2}}{2}}|\tilde{1}\rangle_{a}|\tilde{0}\rangle_{b}
+\sin{\frac{\theta_{0}}{2}}\sin{\frac{\theta_{2}}{2}}|\tilde{1}\rangle_{a}|\tilde{1}\rangle_{b},
\end{align*}
with auxiliary parameters 
\begin{align*}
{\Xi'}_{0}(t)=&\sqrt{\left[\Xi_{\tilde{0}\tilde{0}}(t)\right]^{2}
	+\left[\Xi_{\tilde{0}\tilde{1}}(t)\right]^{2}
	+\left[\Xi_{\tilde{1}\tilde{0}}(t)\right]^{2}
	+\left[\Xi_{\tilde{1}\tilde{1}}(t)\right]^{2}}, \cr   
\theta_{0}=&2\arctan\left[\frac{\sqrt{\Xi^{2}_{\tilde{1}\tilde{0}}(t)+\Xi^{2}_{\tilde{1}\tilde{1}}(t)}}
{{\sqrt{\Xi^{2}_{\tilde{0}\tilde{0}}(t)+\Xi^{2}_{\tilde{0}\tilde{1}}(t)}}}\right], \cr
\theta_{1}=&2\arctan\left[\frac{\Xi_{\tilde{0}\tilde{1}}(t)}{\Xi_{\tilde{0}\tilde{0}}(t)}\right], \ \ \ \theta_{2}=2\arctan\left[\frac{\Xi_{\tilde{1}\tilde{1}}(t)}{\Xi_{\tilde{1}\tilde{0}}(t)}\right].                     
\end{align*}
For simplicity, we choose $\theta_{0,(1,2)}$ to be time-independent.
The orthogonal partners of the state $|b'\rangle$ become
\begin{align*}
|d_{1}\rangle=&
e^{-i\phi}\sin{\frac{\theta_{0}}{2}}\cos{\frac{\theta_{1}}{2}}|\tilde{0}\rangle_{a}|\tilde{0}\rangle_{b}
+\sin{\frac{\theta_{0}}{2}}\sin{\frac{\theta_{1}}{2}}|\tilde{0}\rangle_{a}|\tilde{1}\rangle_{b}\cr
&-\cos{\frac{\theta_{0}}{2}}\cos{\frac{\theta_{2}}{2}}|\tilde{1}\rangle_{a}|\tilde{0}\rangle_{b}
-\cos{\frac{\theta_{0}}{2}}\sin{\frac{\theta_{2}}{2}}|\tilde{1}\rangle_{a}|\tilde{1}\rangle_{b},
\cr
|d_{2}\rangle=&
e^{-i\phi}\cos{\frac{\theta_{0}}{2}}\sin{\frac{\theta_{1}}{2}}|\tilde{0}\rangle_{a}|\tilde{0}\rangle_{b}
-\cos{\frac{\theta_{0}}{2}}\cos{\frac{\theta_{1}}{2}}|\tilde{0}\rangle_{a}|\tilde{1}\rangle_{b}\cr
&+\sin{\frac{\theta_{0}}{2}}\sin{\frac{\theta_{2}}{2}}|\tilde{1}\rangle_{a}|\tilde{0}\rangle_{b}
-\sin{\frac{\theta_{0}}{2}}\cos{\frac{\theta_{2}}{2}}|\tilde{1}\rangle_{a}|\tilde{1}\rangle_{b},
\cr
|d_{3}\rangle=&
e^{-i\phi}\sin{\frac{\theta_{0}}{2}}\sin{\frac{\theta_{1}}{2}}|\tilde{0}\rangle_{a}|\tilde{0}\rangle_{b}
-\sin{\frac{\theta_{0}}{2}}\cos{\frac{\theta_{1}}{2}}|\tilde{0}\rangle_{a}|\tilde{1}\rangle_{b}\cr
&-\cos{\frac{\theta_{0}}{2}}\sin{\frac{\theta_{2}}{2}}|\tilde{1}\rangle_{a}|\tilde{0}\rangle_{b}
+\cos{\frac{\theta_{0}}{2}}\cos{\frac{\theta_{2}}{2}}|\tilde{1}\rangle_{a}|\tilde{1}\rangle_{b}.
\end{align*}

Then, by using the same strategy as that of the single-qubit case, we choose
\begin{align*}
{\Xi'}_{0}(t)\sin{\phi_{\tilde{0}\tilde{0}}}
=&\Omega_{{p}}({\beta},{\varphi})/2
=\dot{{\beta}} \cot {\varphi} \sin {\beta}+\dot{{\varphi}} \cos {\beta},\cr {\Xi'}_{0}(t)\cos{\phi_{\tilde{0}\tilde{0}}}
=&\Omega_{{s}}({\beta'},{\varphi})/2
=\dot{{\beta}} \cot {\varphi} \cos {\beta}-\dot{{\varphi}} \sin {\beta}.
\end{align*}
The evolution operator after a cyclic evolution along 
\begin{align}
|\psi'_{-}(t)\rangle=ie^{i{\beta}}({\varphi}/2)|\mu\rangle|b\rangle+\sin({\varphi}/2)|\zeta'_{{0}}\rangle,
\end{align} 
in the subspace spanned by $\{|b'\rangle,|d_{1}\rangle,|d_{2}\rangle,|d_{3}\rangle\}$ is given by
\begin{align}
U'_{T}=\left(\begin{array}{cccc}
\exp{(2i\Theta_{{s}})} & 0 & 0 & 0\\
0 & 1 & 0 & 0\\
0 & 0 & 1 & 0\\
0 & 0 & 0 & 1
\end{array}
\right).
\end{align}
In the computational subspace spanned by $\{|\tilde{0}\rangle_{a}|\tilde{0}\rangle_{b},~|\tilde{0}\rangle_{a}|\tilde{1}\rangle_{b},~|\tilde{1}\rangle_{a}|\tilde{0}\rangle_{b},~|\tilde{1}\rangle_{a}|\tilde{1}\rangle_{b}\}$, the evolution operator $U'_{T}$ 
	describes a universal two-qubit geometric gate (see table~\ref{tab2} for examples). 
	These two-qubit gates using the same strategy as the single-qubit case are also insensitive
	to the errors induced by pulse imperfections. 
	Therefore, when considering the error coefficient $\delta_{{i}}=0.1$,
	in Fig.~\ref{fig5}(b),
	we show that arbitrary two-qubit gates can be implemented with high fidelities.

}

\section{Preparing superpositions of Fock states}
	
\begin{figure}
	\centering
	\scalebox{0.75}{\includegraphics{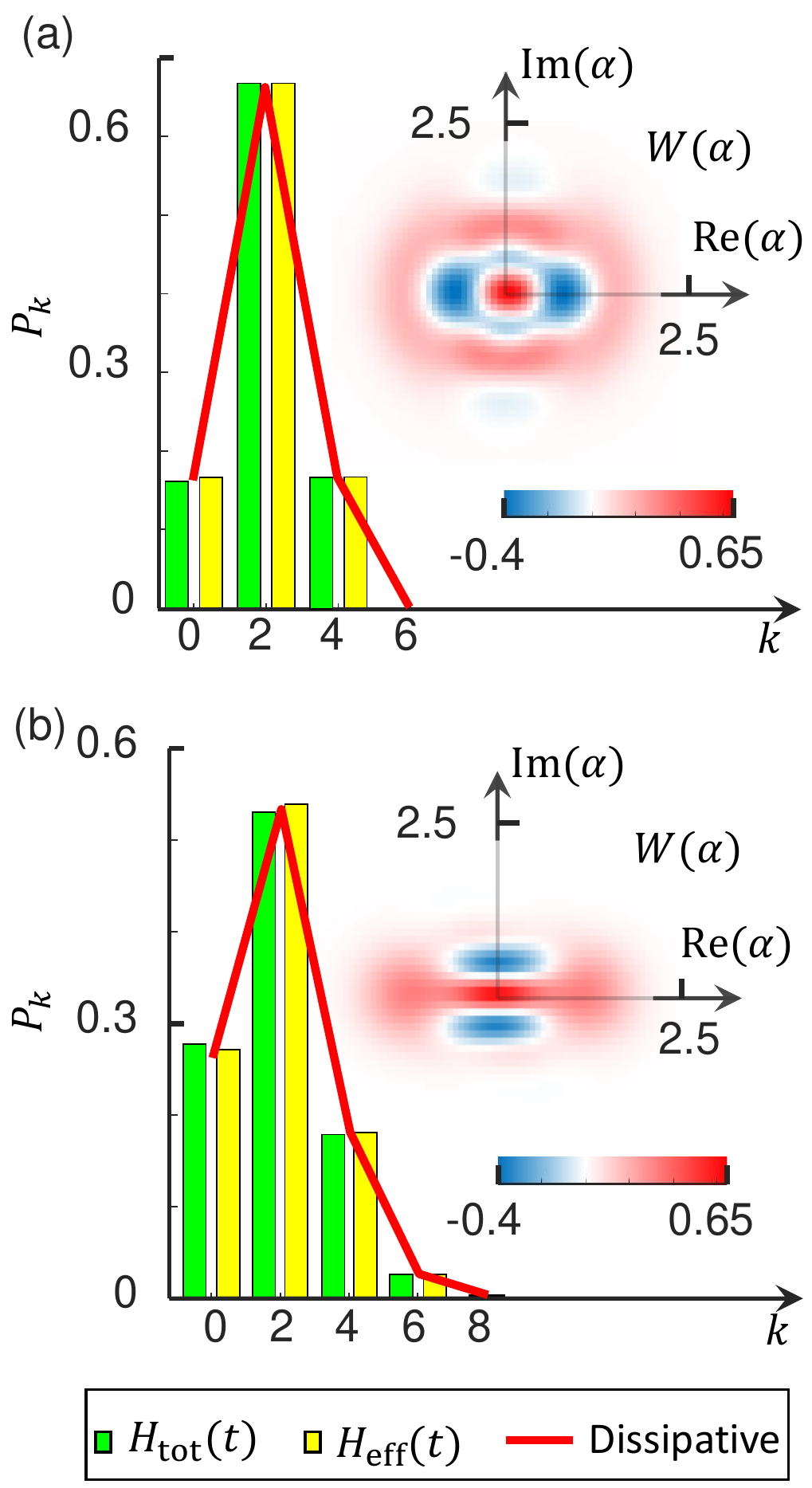}}
	\caption{Histograms of Fock-state populations calculated with the total Hamiltonian $H_{\rm{tot}}(t)$ (green) and the effective Hamiltonian $H_{\rm{eff}}(t)$ (yellow). The red-solid broken lines in each panel denotes
		the Fock-state population calculated by the master equation in Eq.~(\ref{eq1-6}) when $\left(\kappa,\kappa^{\phi},\gamma_{g,(\mu)},\gamma_{g,(\mu)}^{\phi}\right)\simeq2\pi\times\left(0.33,0.3,8,8\right)~{\rm{kHz}}$ \cite{Prl124120501}.
		Insets show the Wigner
		function $W(\alpha)$ of the generated states. (a) The superposition state $|\psi_{\rm{in}}\rangle=(|\tilde{0}\rangle+\sqrt{2}|\tilde{1}\rangle)/\sqrt{3}$ with parameters
		$\tilde{\beta}_{f}=\arccos(\sqrt{1/6})$,  $\epsilon_{2}=2/\sqrt{5}$, and $g=0.7\omega_{{c}}$.
		(b) The cat state $|\mathcal{C}_{{e}}^{\eta}\rangle$ with amplitude
		$\eta=g/\omega_{{c}}=\sqrt{2}$.
		The evolution time for each panel is
		$T=35~{\rm{ns}}$.
	}
	\label{fig1}
\end{figure}

High-fidelity input states are needed
to verify the feasibility of the proposed NHQC in experiments.
To generate these input states, starting from Eq.~(\ref{eq1-2}), we assume 
\begin{align}
 \phi_{0}&=0,\ \ \ \ \ \
 c_{k'}^{m}\Omega_{k'}=\epsilon_{k'}\Omega_{{s}}(\tilde{\beta},\tilde{\varphi}), \cr
 \phi_{k'}&=\pi, \ \  \ \ \ \
 c_{0}^{m}\Omega_{0}=\Omega_{{p}}(\tilde{\beta},\tilde{\varphi})
\end{align}
where,
$k'\neq 0$ are even numbers,
$\epsilon_{k'}$ are time-independent coefficients satisfying $\sum_{k'}|\epsilon_{k'}|^2=1$, $\tilde{\beta}$ and $\tilde{\varphi}$
satisfying $\tilde\beta(0)\simeq\tilde\varphi(0)\simeq0$ are time-dependent
auxiliary parameters to be determined.
Then, the evolution governed by $H_{\rm{eff}}(t)$ is
\begin{align}
|{\tilde\psi_{0}(t)}\rangle=&\cos{\tilde\varphi}(\cos{\tilde\beta}|0\rangle+\sin{\tilde\beta}\sum\nolimits_{k'}\epsilon_{k'}|k'\rangle)|\mu\rangle\cr &-i\sin{\tilde\varphi}|\zeta_{{2}}\rangle.
\end{align}
When $\tilde\varphi(t_{f})\simeq0$ and $\tilde\beta(t_{f})=\tilde\beta_{f}$, we obtain \begin{align}|\tilde\psi_{0}(t_{f})\rangle=\left(\cos{\tilde\beta_{f}}|0\rangle+\sin{\tilde\beta_{f}}\sum\nolimits_{k'}\epsilon_{k'}|k'\rangle\right)|\mu\rangle,\end{align}
which is an arbitrary superposition of even-number Fock states.
The boundaries for $\tilde\beta$ and $\tilde\varphi$ can be satisfied by choosing
\begin{align}
  \tilde\beta&=\frac{\tilde\beta_{f}}{\left[1+\exp{\left({-{t}/{\tau}+{T}/{2\tau}}\right)}\right]},\cr
  \tilde\varphi&=\frac{\tilde\varphi_{0}} {\exp{\left({t}/{\tau_{c}}-{T}/{2\tau_{c}}\right)^{2}}},
\end{align}
with parameters $\left(\tilde\varphi_{0},\tau,\tau_{c}\right)=\left(\pi/5,0.11
5T,0.3T\right)$ \cite{Pra80013417,Rmp701003}.

For instance, when $k=(0,2,4)$, $m={2}$, $\tilde{\beta}_{f}=\arccos(1/\sqrt{6})$, and $\epsilon_{2}=2/\sqrt{5}$
we can generate an input state $|\psi_{\rm{in}}\rangle=(|\tilde{0}\rangle+\sqrt{2}|\tilde{1}\rangle)/\sqrt{3}$,
as shown in Fig.~\ref{fig1}(a).
This figure shows the final populations $P_{k}=\langle\mu|\langle k|\rho(t_{{f}})|k\rangle|\mu\rangle$
and the Wigner function $W(\alpha)={2}{\rm{Tr}}[D_{\alpha}^{\dag}\rho(t_{{f}})D_{\alpha}e^{i\pi a^{\dag}a}]/\pi$,
where $D_{\alpha}=\exp(\alpha a^{\dag}-\alpha^{*}a)$ is the displacement operator.
As shown in Fig.~\ref{fig1}(a) the full dynamics [green histograms] is in excellent agreement with the
effective dynamics [yellow histograms].

In the presence of decoherence, the populations [red-solid broken line in Fig.~\ref{fig1}(a)], calculated using the master equation
in Eq.~(\ref{eq1-6})
are almost the same as those calculated using the coherent dynamics when feasible parameters are considered. %$\left(\kappa,\kappa^{\phi},\gamma_{g,(\mu)},\gamma_{g,(\mu)}^{\phi}\right)\simeq\left(2.1,1.8,28,40\right){\rm{kHz}}$.
This indicates
that our protocol for the state preparations is robust against decoherence.

The above approach can be used to generate Schr\"{o}dinger's cat states \cite{Phys72597,Pra71063820,Np7799,Pra100012124,Zhou2021,Qin2021}, e.g., the even cat state 
\begin{align}
|\mathcal{C}_{{e}}^{\eta}\rangle=e^{|\eta|^2/2}\sqrt{\rm{sech}|\eta|^2}(|\eta\rangle+|-\eta\rangle)/2,
\end{align} 
when $m={0}$,
  ${\tilde{\beta}_{f}}=\arccos{\left(\sqrt{{\rm{sech}|\eta|^2}}\right)}$ and $\epsilon_{k'}=-(\eta^{k'}\cot{\tilde{\beta}_{f}})/\sqrt{k'!}$,
where $\eta$ is the amplitude of the coherent state $|\eta\rangle$.
In Fig.~\ref{fig1}(b), we show that the even cat state can be generated with a high fidelity.
These generated high-fidelity cat stats are useful for cat-code quantum computation \cite{Prl116140502,Njp16045014}.

%The cat state can be generated with a high fidelity as shown in %Fig.~\ref{fig1}(b).

%via the binomial
%codes %$\{|\tilde{0}\rangle_{a}|\tilde{0}\rangle_{b},|\tilde{0}\rangle_{a}|\tilde{1}\rangle_{b},|\tilde{1}\rangle_{a}|\t%ilde{0}\rangle_{b},|\tilde{1}\rangle_{a}|\tilde{1}\rangle_{b}\}$.

\section{Conclusion}

We have investigated the possibility of using
USC systems for the implementation of \textit{fast, robust, and fault-tolerant} holonomic computation.
The dressed-state properties of the USC systems allow
to \textit{simultaneously couple} the dressed state $|\zeta_{m}\rangle$ to multiple Fock sates,
such that one can manipulate the population and the phase of each Fock state as desired.
The binomial codes formed from these Fock states are protected against the single-photon loss, making the computation fault-tolerant.
Moreover, by designing the pulses with invariant-based engineering,
we can eliminate the dynamical phase and achieve only the geometric phase
in a cyclic evolution.
Such a control technique is compatible with the systematic-error-sensitivity nullification
method, making the evolution mostly insensitive to the systematic errors caused by pulse imperfections.
Additionally, using the USC regime allows to apply relatively strong driving fields, such that our protocols are
fast. As results, our protocols are robust against the
decays and dephasings of the cavity and the atom.
Note that this work can freely control a bosonic mode. The proposed idea can be generalized to realize
NHQC with other bosonic error-correction qubits, such as cat-qubits \cite{Prl116140502,Njp16045014}, for
fast, robust, and fault-tolerant quantum computation. 
%{Thus, our protocol
%is promising for building blocks for a surface code tailored to biased noise with high error thresholds %\cite{Prl120050505,Prx9041031,Prl124130501,Qst5043001}.}

The proposed protocols can be realized in superconducting circuits
\cite{Nr119,Rmp91025005,Rmp841,Rmp85623,Pr7181,Nat474589,An16767,Pra90053833,Prl103147003,Np6772,Prl105237001,npjQI346,Np1344,Sr626720,Prb93214501,Pra96012325,Pra95053824,Prl120183601}. For instance, one can inductively
couple a flux qubit and an $LC$ oscillator via Josephson
junctions \cite{Np1344} to reach the needed coupling strength.
%light-matter coupling strength $g/\omega_{{c}}\sim 1.34$ has been reached in experiments
The quantized level structure in Fig.~\ref{figmodel}(b) can be realized adjusting
the external magnetic flux through the qubit loop \cite{Prl110243601,Njp19053010,Pra89033827}.

{Some experimental observations of the ultrastrong
light-matter coupling in superconducting quantum circuits are listed in Table~\ref{tabS2}.
To reach the ultrastrong and deep-strong coupling regimes, we can choose 
a setup with a flux qubit coupled to a lumped-element $LC$ resonator
\cite{Np1344,Pra95053824}.
In such superconducting circuit experiments, qubit and resonator frequencies are
usually in the range $\omega_{c,(q)}/2\pi\sim1$--$10~{\rm{GHz}}$. Thus, we choose
$g/\omega_{c}\simeq 0.7~(0.8)$ and $\omega_{c}/2\pi=6.25~{\rm{GHz}}$, which are experimentally feasible, as shown in Table~\ref{tabS2}. }

{\renewcommand\arraystretch{1.4}
\begin{table}
			\centering
			\caption{Superconducting experiments that have achieved the ultrastrong light-matter coupling. Abbreviations are FQ=flux qubit, TR=transmon qubit, TL=transmission line resonator, and LE=lumped-element resonator.}
			\label{tabS2}
			\begin{tabular}{p{2cm}<{\centering}p{1cm}<{\centering}p{1cm}<{\centering}p{1cm}<{\centering}p{1cm}<{\centering}p{1cm}<{\centering}}
				\hline
				\hline
				{Year \& Ref.}  & {Qubit} & Cavity & $g/2\pi$ (MHz) & $\omega_{c}/2\pi$ (GHz) & $g/\omega_{c}$ \\
				\hline
				2010 \cite{Np6772} & FQ & TL & 636 & 5.357 &  0.12 \\
				2010 \cite{Prl105237001} & FQ &LE& 810 & 8.13& 0.1 \\
				2017 \cite{Np1344} & FQ &LE & 7630 & 5.711 & 1.34 \\
				2017 \cite{Pra95053824} &FQ &LE &  5310 & 6.203 & 0.86 \\
				2017 \cite{npjQI346} &TR & TL &  897 & 4.268 & 0.19 \\
				2018 \cite{Prl120183601} & FQ & LE & 7480 & 6.335 & 1.18\\
				\hline
				\hline
			\end{tabular}
\end{table}}

Recent experimental work has demonstrated
that dissipation and dephasing rates in a flux qubit is of the order of
$2\pi\times 10~{\rm{kHz}}$  \cite{Rmp91025005,Prl113123601,Prb93104518}.
The transmon qubits, which have lower anharmonicity than flux qubits, can have dissipation and dephasing rates approaching $2\pi\times 1~\rm{kHz}$ \cite{Prb75140515,Nc712964}.
For transmission-line resonators, quality factors  factors $Q=\omega_{c}/\kappa$ on the order of $10^{6}$ have been realized \cite{Prl107240501}, which indicates that quantum coherence of single photons up to $1 \sim 10 ~\rm{ms}$ is within current experimental capabilities \cite{Apl100113510}. 
Therefore, our proposal works well in the
USC regime, and it may find compelling applications
for quantum information processing for various
USC systems, in particular, superconducting systems.

\begin{acknowledgements}
	We acknowledge helpful discussions with Y.-H. Kang and Z.-B. Yang.
	Y.-H.C. is supported by the Japan Society for the Promotion of Science (JSPS) KAKENHI Grant No.~JP19F19028.
	X.W. is supported by the China
	Postdoctoral Science Foundation No.~2018M631136,
	and the Natural Science Foundation of China under
	Grant No.~11804270.
	 F.N. is supported in part by:
	 Nippon Telegraph and Telephone Corporation (NTT) Research,
	 the Japan Science and Technology Agency (JST) [via
	 the Quantum Leap Flagship Program (Q-LEAP),
	 the Moonshot R\&D Grant No.~JPMJMS2061, and
	 the Centers of Research Excellence in Science and Technology (CREST) Grant No.~JPMJCR1676],
	 the Japan Society for the Promotion of Science (JSPS)
	 [via the Grants-in-Aid for Scientific Research (KAKENHI) Grant No.~JP20H00134 and the
	 JSPS–RFBR Grant No. JPJSBP120194828],
	 the Army Research Office (ARO) (Grant No.~W911NF-18-1-0358),
	 the Asian Office of Aerospace Research and Development (AOARD) (via Grant No.~FA2386-20-1-4069), and
	 the Foundational Questions Institute Fund (FQXi) via Grant No.~FQXi-IAF19-06.
\end{acknowledgements}

\begin{appendix}

	\section{Effective Hamiltonian}\label{AA}
	The total Hamiltonian for this protocol can be written as
	\begin{align}\label{eqs1-1}
	H_{\rm{tot}}&=H_{0}+H_{{D}}(t), \cr
	H_{0}&=\hbar\sum_{m=0}^{\infty}E_{m}|\zeta_{m}\rangle\langle\zeta_{m}|
	+\sum_{n=0}^{\infty}\hbar(\omega_{\mu}+n{\omega}_{{c}})|\mu\rangle\langle\mu|\otimes|n\rangle\langle n|, \cr
	H_{{D}}(t)&=\hbar\Omega(|\mu\rangle\langle g|+|g\rangle\langle\mu|).
	\end{align}
	Here, $|\zeta_{m}\rangle$ are the dressed eigenstates of the Rabi Hamiltonian with eigenvalues $E_{m}$,
	$\omega_{\mu}$ denotes the energy of the lowest atomic level $|\mu\rangle$, $n$ is the cavity photon number, and
	$\Omega=\Omega_{k}\cos(\omega_{k}t+\phi_{k})$ 
	is a composite pulse driving the atomic transition $|\mu\rangle\leftrightarrow|g\rangle$.
	Performing the unitary transformation $U_{d}=\exp(-iH_{0}t/\hbar)$ and choosing the frequencies as 
	$\omega_{{k}}=E_{m}-\omega_{\mu}-k\omega_{{c}}$,  
	we have  
	\begin{widetext}
	\begin{align}\label{eqs1-2}
	H'_{D}(t)=&\frac{\hbar}{2}\sum_{k}\sum_{m'}\sum_{n}c_{n}^{m'}{\Omega_{k}}|\mu\rangle|n\rangle\langle\zeta_{m'}|\left\{\right.\exp{\left[-i\Delta E_{m,m'}t+i (n-k) {\omega}_{{c}}t+i\phi_{k}\right]}\cr
	&+\exp{[-i\Delta E_{m,m'}t+i (n-k) {\omega}_{{c}}t-2i\omega_{{k}}t-i\phi_{k}]}\left.\right\}
	+\rm{H.c.},
	\end{align}
	\end{widetext}
	where $\Delta E_{m,m'}=E_{m'}-E_{m}$ is the energy gap between the eigenstates $|\zeta_{m'}\rangle$ and $|\zeta_{m}\rangle$.

\begin{figure*}
	\centering
	\scalebox{0.43}{\includegraphics{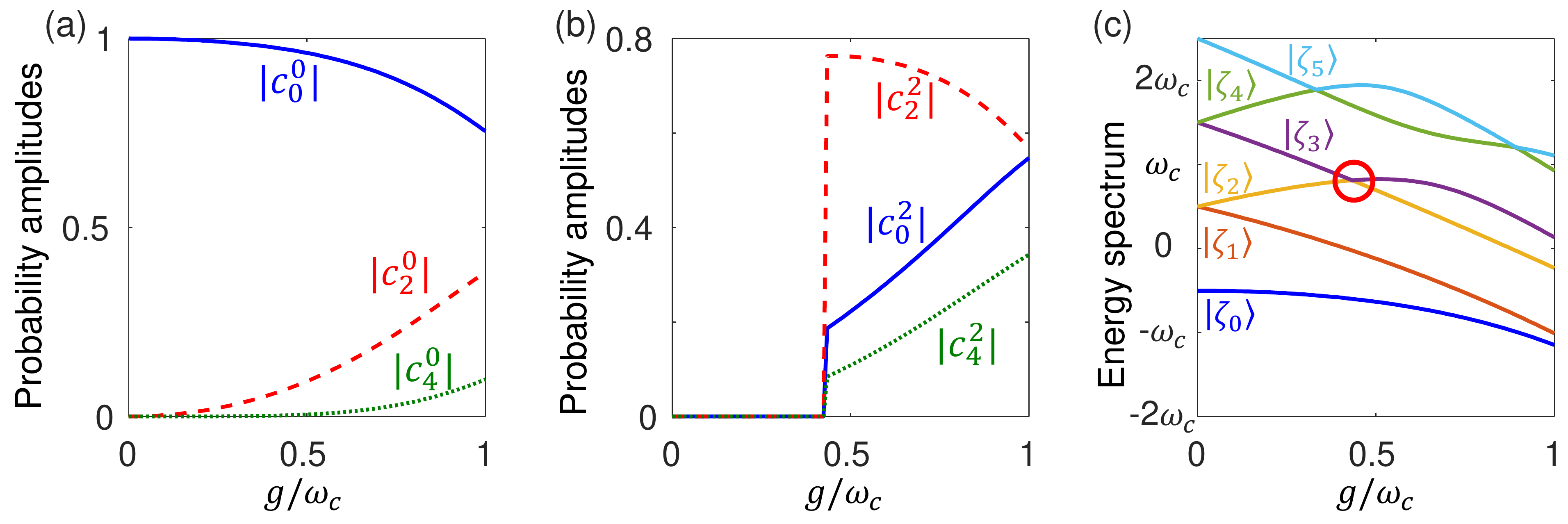}}
	\caption{{Probability amplitudes of $|g\rangle|0\rangle$, $|g\rangle|2\rangle$, and $|g\rangle|4\rangle$ in dressed states (a) $|\zeta_{{0}}\rangle$ and (b) $|\zeta_{{2}}\rangle$ of $H_{R}$ as a function of the coupling strength $g$ when $\omega_{q}=\omega_{c}$. (c) Energy spectrum of the Hamiltonian $H_{R}$ when $\omega_{q}=\omega_{c}$. The red circle in panel (c) denotes an avoided level crossing.
	}}
	\label{figS2}
\end{figure*}

	%\begin{figure}[b]
	%\centering
	%\scalebox{0.5}{\includegraphics{c00c02c04.pdf}}
	%\caption{Probability amplitudes of $|g\rangle|0\rangle$, $|g\rangle|2\rangle$, and $|g\rangle|4\rangle$ in the ground %state $|\zeta_{{0}}\rangle$ of $H_{\rm{R}}$ as a function of the coupling strength%
	%	$g$ for the resonant case $\omega_{\rm{q}}=\omega_{\rm{c}}$.}%
	%	\label{figS0}
	%\end{figure}

	Obviously,
	when satisfying
	\begin{align}\label{eqS5}
	c_{n}^{m'}\Omega_{k}\ll ~ &|(n-k)\omega_{{c}}-\Delta E_{m,m'}| , \cr
	c_{n}^{m'}\Omega_{k}\ll ~ &|(n-k)\omega_{{c}}-2\omega_{k}-\Delta E_{m,m'}|, 
	\end{align}
	the fast-oscillating terms can be neglected 
	in the rotating wave approximation (RWA).
	Then, the effective Hamiltonian becomes
	\begin{align}\label{eqs1-4}
	H_{\rm{eff}}(t)=&\frac{\hbar}{2}\sum_{k}{c_{k}^{m}\Omega_{k}}e^{i\phi_{k}}|\mu\rangle|k\rangle\langle\zeta_{m}|+{\rm{H.c.}},
	\end{align}
	i.e., the effective Hamiltonian in Eq.~(\ref{eq1-2}). 
		The coefficients $c_{n}^{m}=\langle\zeta_{m}|g\rangle|n\rangle$ and $d_{n\pm 1}^{m}=\langle\zeta_{m}|e\rangle|n\pm 1\rangle$
		can be obtained numerically [see Fig.~\ref{figS2}(a) as an example for
		the ground dressed state $|\zeta_{{0}}\rangle$]. 
		According to our numerical results, 
		when $0.5\lesssim g/\omega_{c}\lesssim 1$, the probability amplitudes $\left(c_{0}^{{2}},c_{2}^{{2}},c_{4}^{{2}}\right)$ of the states $\left(|g\rangle|0\rangle,|g\rangle|2\rangle,|g\rangle|4\rangle\right)$
		in the third dressed state $|\zeta_{{2}}\rangle$ are greater [see Fig.~\ref{figS2}(b)] than in the others.
		Thus, when focusing on manipulating the Fock states ($|0\rangle,|2\rangle,|4\rangle$),
		the effective driving intensities (i.e., $c_{0}^{{2}}\Omega_{0}$, $c_{2}^{{2}}\Omega_{2}$, and $c_{4}^{{2}}\Omega_{4}$) can be much stronger by using
		$|\zeta_{{2}}\rangle$ to be the intermediate state. 
		Therefore, the gate time can be shortened. 
		In Fig.~\ref{figS2}(b), we find that the coefficients $c_{n}^{m}$ jump from zero to 
		nonzero values when $g/\omega_{c}\simeq 0.43$. This is caused by an avoided level crossing \cite{Nr119,Rmp91025005} when
		$g/\omega_{c}\simeq 0.43$ [see the red circle in Fig.~\ref{figS2}(c)].
		In Fig.~\ref{figS2}(c) we notice that 
		the dressed states become nonequidistant as the energy gap $\Delta E_{m,m+1}\neq {\rm{constant}}$ when $0.1\lesssim g/ \omega_{c}\lesssim1$.
		For instance, when $g/\omega_{c}\sim 0.5$,
		we have $|\Delta E_{m,m+1}-\Delta E_{m+1,m+2}|\gtrsim 0.5\omega_{c}$.
		This indicates that the USC can induce 
		strong anharmonicity in the dressed states $|\zeta_{m}\rangle$.

		%$g/\omega_{c}\ll 1$ (in the weak coupling regime), we obtain ($l=0,1,2,\ldots$)
		%\begin{eqnarray}
		%|c_{0}^{{0}}|=&1, \ \ \ \ \ \ \ \ \ \ \ c_{n\neq 0}^{{0}}&=0, \cr\cr
		%|c_{2l+1}^{2l+1}|=&1/\sqrt{2}, \ \ \ c_{n\neq 2l+1}^{2l+1}&=0, \cr \cr
		%|c_{2l+1}^{2l+2}|=&1/\sqrt{2}, \ \ \ c_{n\neq 2l+1}^{2l+2}&=0. 
		%\end{eqnarray}
		%This means that when in the weak coupling regime, the effective Hamiltonian
		%$H_{\rm{eff}}(t)$ becomes
		%\begin{align}
		%  H_{\rm{eff}}(t)=&\frac{\Omega_{0}}{2}\exp{(i\phi_{0})}|\mu\rangle|0\rangle\langle\zeta_{{0}}|+{\rm{H.c.}}, \ \ \ \ \ \  ({\text{when}\ \ m={0}}), \cr\cr
		%  H_{\rm{eff}}(t)=&\frac{\Omega_{0}}{2}\exp{(i\phi_{0})}|\mu\rangle|0\rangle\langle\zeta_{{0}}|+{\rm{H.c.}}, \ \ \ \ \ \  ({\text{when}\ \ m={0}}), \cr\cr
		%  H_{\rm{eff}}(t)=&\frac{\Omega_{0}}{2}\exp{(i\phi_{0})}|\mu\rangle|0\rangle\langle\zeta_{{0}}|+{\rm{H.c.}}, \ \ \ \ \ \  ({\text{when}\ \ m={0}}),
		%\end{align}

	%Note that when entering the deep-strong coupling (DSC) regime,
	%it is possible that $k\omega_{\rm{c}}-\Delta E_{2k}=0$ while $c_{2k}^{k}\neq0$ ($k\neq0$) for some specific
	%coupling strengths (see Tab.~\ref{tab1}).
	%We need to avoid using such coupling strengths,
	%which can make the conditions in Eq.~(\ref{eqS5}) unsatisfied.
	%Figure~\ref{figS0} shows the probability amplitudes $|c_{0}^{0}|$, $|c_{0}^{2}|$, and $|c_{0}^{4}|$.
	%We can find that $|c_{0}^{4}|$ becomes significant when the coupling enters
	%the DSC regime, e.g., $g/\omega_{{c}}\sim 1.0$. Hence, to generate and manipulate the
	%Fock states $|k\geq4\rangle$ with high fidelities, we can choose $g/\omega_{{c}}\sim 1.0$
	%for our protocols. 

	\section{Dynamical and geometric phases}\label{AB}
	An operator $I(t)$ satisfying $\hbar{\partial_{t}}I(t)=i[H(t),I(t)]$ is a dynamical invariant of an arbitrary Hamiltonian $H(t)$. According to \cite{Jmp101458}, an arbitrary solution of the Schr\"{o}dinger equation 
	\begin{align}\label{eq2s-5}
	i\hbar\frac{\partial}{\partial t}|\psi(t)\rangle=H(t)|\psi(t)\rangle,
	\end{align}
	can be expressed by using the eigenstates of $I(t)$ as 
	\begin{align}\label{eq2s-6}
	|\psi(t)\rangle=&\sum_{n} {C_{n}e^{i\mathcal{R}_{n}(t)}|{\psi_{n}(t)}}\rangle, \cr
	\mathcal{R}_{n}(t)=&\frac{1}{\hbar}\int_{0}^{t}\langle{\psi_{n}(t')}|\left[i{\hbar}{\partial_{t'}}-{H(t')}\right]|{\psi_{n}(t')}\rangle dt',
	\end{align}
	where $C_{n}$ are time-independent amplitudes, $|\psi_{n}(t)\rangle$ are the orthonormal eigenvectors of $I(t)$, and $\mathcal{R}_{n}(t)$
	are the Lewis-Riesenfeld phases \cite{Jmp101458}. 
	These phases include dynamical phases 
	\begin{align}
	\vartheta_{n}(t)=-\frac{1}{\hbar}\int_{0}^{t}\langle{\psi_{n}(t')}|H(t')|{\psi_{n}(t')}\rangle dt', 
	\end{align}
	and geometric phases
	\begin{align}
	\Theta_{n}(t)=\int_{0}^{t}\langle{\psi_{n}(t')}|{i\partial_{t'}}|{\psi_{n}(t')}\rangle dt'.
	\end{align}
	For instance, when $\langle\psi(0)|\psi_{0}(0)\rangle=1$, we have $C_{0}=1$ and $C_{n\neq0}=0$.
		The evolution of the system is exactly along the eigenstate $|\psi_{0}(t)\rangle$, which
		is a shortcut to the adiabatic passage of $H(t)$.
	%By imposing $[I(0),H(0)]=[I(t_{{f}}),H(t_{{f}})]=0$, we have $|{\psi_{n}(0)}\rangle=|{n(0)}\rangle$ and $|{\psi_{n}(t_{{f}})}\rangle=|{n(t_{{f}})}\rangle$, 
	%where $|n(t)\rangle$ are the eigenstates of the Hamiltonian $H(t)$ and $t_{{f}}$ is the final time. 
	%Therefore, the eigenvectors $|\psi_{n}(t)\rangle$ can be chosen to be the shortcuts-to-adiabatic (STA) passages
	%for the generations of Fock-state superpositions and the nonadiabatic Holonomic gates.

	The effective Hamiltonian 
	\begin{align}\label{eqS9}
	{H}_{\rm{eff}}(t)=\frac{\hbar}{2}{\Omega}_{0}e^{i{\phi}_{2}}|\zeta_{{2}}\rangle\langle b|\langle\mu|+{\rm{H.c.}},
	\end{align} 
	in Eq.~(\ref{eq9}) for the NHQC can be regarded 
	as the intermediate state $|\zeta_{{2}}\rangle$
	coupled to the bright state 
	\begin{align}\label{eqS10}  
	|b\rangle=e^{-i\phi}\sin({\theta}/{2})|\tilde{0}\rangle+\cos({\theta}/{2})|\tilde{1}\rangle,
	\end{align} 
	but decoupled from the dark state 
	\begin{align}\label{eqS11}
	|d\rangle=e^{-i\phi}\cos({\theta}/{2})|\tilde{1}\rangle-\sin({\theta}/{2})|\tilde{0}\rangle.
	\end{align}
	A dynamical invariant of ${H}_{\rm{eff}}(t)$ is 
	\begin{align}\label{eqS12}
	{I}(t)=&\cos{{\varphi}}(|\zeta_{{2}}\rangle\langle\zeta_{{2}}|
	-|b\rangle\langle b|\otimes|\mu\rangle\langle \mu|)\cr
	&+\left(e^{i{\beta}}\sin{{\varphi}}|\zeta_{{2}}\rangle\langle b|\langle\mu|+{\rm{H.c.}}\right),
	\end{align}
	with eigenvectors
	\begin{align}\label{eqS13}
	|\psi_{+}(t)\rangle&=\sin({\varphi}/2)|\mu\rangle|b\rangle+ie^{-i{\beta}}\cos({\varphi}/2)|\zeta_{{2}}\rangle, \cr
	|\psi_{-}(t)\rangle&=ie^{i{\beta}}\cos({\varphi}/2)|\mu\rangle|b\rangle+\sin({\varphi}/2)|\zeta_{{2}}\rangle.
	\end{align}
	
	Then, substituting Eqs.~(\ref{eqS9}) and (\ref{eqS13}), 
	into Eq.~(\ref{eq2s-6}), the time derivatives of
	the dynamic phases and geometric phases acquired 
	by $|{\psi}_{\pm}(t)\rangle$ are 
	\begin{align}\label{eqS14}
	\dot{\vartheta}_{\pm}(t)=&\mp\frac{\dot{{\beta}}}{2}\sin{\varphi}\tan{{\varphi}}, \cr
	\dot{\Theta}_{\pm}(t)=&\pm\frac{\dot{{\beta}}}{2}(1-\cos{\varphi}),
	\end{align}
	respectively.
	Obviously, $\dot{\vartheta}_{\pm}(t)$ and 
	$\dot{\Theta}_{\pm}(t)$ obey the same mathematical symmetry.
	To eliminate the dynamical phases and
	achieve only the geometric phases, we can
	design a piecewise function for ${\beta}$, e.g.,
	\begin{align}\label{eqS15}
	{\beta}=\left\{\begin{array}{ll}
	f(t),& t\in[0,t_{{f}}/2] \\ \\
	f(t)-2\Theta_{{s}},& t\in[t_{{f}}/2,t_{{f}}]
	\end{array}
	\right.
	\end{align}
	where $\Theta_{{s}}$ is a constant.
	Then, we assume $\dot{\vartheta}_{\pm}(t-t_{{f}}/2)$ to be odd functions, leading to
	\begin{align}\label{eqS16}
	\vartheta_{\pm}&=\mp\int_{\frac{t_{{f}}}{2}}^{\frac{t_{{f}}}{2}+\Delta t}\Theta_{{s}}(\sin{\varphi}\tan{{\varphi}})dt+\int_{0}^{t_{{f}}}\dot{\vartheta}_{\pm}(t) dt \cr
	&=\mp\Theta_{{s}}\sin{\varphi}\left(\frac{t_{{f}}}{2}\right)\tan{{\varphi}\left(\frac{t_{{f}}}{2}\right)}.
	\end{align}
	Here, $\Delta t$ is a small increase in time, and we have assumed ${\varphi}$ to be continuous in time.
	Meanwhile, for the geometric phases, $\dot{\Theta}_{\pm}(t-t_{{f}}/2)$ are also odd functions, leading to
	\begin{align}\label{eqS17}
	\Theta_{\pm}=& \mp\int_{\frac{t_{{f}}}{2}}^{\frac{t_{{f}}}{2}+\Delta t}\Theta_{{s}}(1-\cos{\varphi})dt+\int_{0}^{t_{{f}}}\dot{\Theta}_{\pm}(t) dt \cr
	=&\mp \Theta_{{s}}\left[1-\cos{{{\varphi}\left(\frac{t_{{f}}}{2}\right)}}\right].
	\end{align}
	Thus, we obtain $\vartheta_{\pm}=0$ and $\Theta_{\pm}=\mp2\Theta_{{s}}$ when ${\varphi}(t_{{f}}/2)=\pi$.

\end{appendix}

\bibliography{references}

\end{document}